\documentclass[a4paper,fleqn,usenatbib]{mnras}
\usepackage[T1]{fontenc}
\usepackage{ae,aecompl}

\usepackage{graphicx}	
\usepackage{amsmath}	
\usepackage{amssymb}	
\usepackage{graphics}	
\usepackage{booktabs}   
\usepackage{comment}
\usepackage{caption}
\usepackage{comment}
\usepackage{hyperref}
\usepackage{bm}
\usepackage{subcaption}
\usepackage{amssymb}
\usepackage{pifont}
\title[The Three Hundred: Phase-space]{The Three Hundred Project: Ram pressure and gas content of haloes and subhaloes in the phase-space plane}

\author[J. Arthur et al.]{
Jake Arthur,$^{1}$\thanks{E-mail: jake.arthur@nottingham.ac.uk}
Frazer R. Pearce,$^{1}$
Meghan E. Gray,$^{1}$
Alexander Knebe,$^{2,3,4}$
\newauthor{
Weiguang Cui,$^{2,5}$
Pascal J. Elahi,$^{4}$
Chris Power,$^{4}$
Gustavo Yepes,$^{2,3}$
}
\newauthor{
Alexander Arth,$^{6,7}$
Marco De Petris,$^{8,9}$
Klaus Dolag,$^{6,10}$
Lilian Garratt-Smithson,$^{4}$
}
\newauthor{
Lyndsay J. Old,$^{11,12}$
Elena Rasia,$^{13}$
Adam R. H. Stevens$^{4}$
}
}

\date{Accepted 2019 January 17. Received 2019 January 17; in original form 2018 October 26}

\pubyear{2018}

\newcommand{\Fig}[1]{Fig.~\ref{#1}}
\newcommand{\Sec}[1]{Section~\ref{#1}}
\newcommand{\Eq}[1]{Eq.~\ref{#1}}
\def \ahf{{\sc ahf}}

\def \gadgettwo{{\sc Gadget2}}
\def \gadgetthree{{\sc Gadget3}}
\def \gx{{\sc GadgetX}}
\def \threehun{{\sc TheThreeHundred}}

\begin{document}
\label{firstpage}
\pagerange{\pageref{firstpage}--\pageref{lastpage}}
\maketitle

\begin{abstract}
We use \threehun\ project, a suite of 324 resimulated massive galaxy clusters embedded in a broad range of environments, to investigate (i) how the gas content of surrounding haloes correlates with phase-space position at $z=0$, and (ii) to investigate the role that ram pressure plays in this correlation. By stacking all 324 normalised phase-space planes containing 169287 haloes and subhaloes, we show that the halo gas content is tightly correlated with phase-space position. At $\sim\,1.5-2\,\text{R}_{\text{200}}$ of the cluster dark matter halo, we find an extremely steep decline in the halo gas content of infalling haloes and subhaloes irrespective of cluster mass, possibly indicating the presence of an accretion shock. We also find that subhaloes are particularly gas-poor, even in the cluster outskirts, which could indicate active regions of ongoing pre-processing. By modelling the instantaneous ram pressure experienced by each halo and subhalo at $z=0$, we show that the ram pressure intensity is also well correlated with phase-space position, which is again irrespective of cluster mass. In fact, we show that regions in the phase-space plane with high differential velocity between a halo or subhalo and its local gas environment, are almost mutually exclusive with high halo gas content regions. This suggests a causal link between the gas content of objects and the instantaneous ram pressure they experience, where the dominant factor is the differential velocity.
\end{abstract}




\section{Introduction}


In the $\Lambda$CDM paradigm dense structures, such as galaxy clusters, grow hierarchically via gravitational attraction \citep{white1978}. As galaxies fall into and build these objects, several environmental mechanisms are thought to significantly alter the galaxies properties \citep[for review see][]{boselli2006}.

Numerous observational studies in this area agree that ram pressure stripping \citep[RPS,][]{gunn1972} is particularly effective at shaping final galaxy properties in clusters. For example, recent observations have revealed H\,{\sc i} (neutral, atomic Hydrogen) gas trailing from galaxies that are entering cluster environments \citep{haynes1984, cayatte1990, jaffe2015, gavazzi2018}. H\,{\sc i} is considered to fuel star formation \citep{prochaska2009,vandevoort2012} and hence is an important factor in galaxy evolution. Jellyfish galaxies are examples of this stripping of H\,{\sc i} in action; here the trailing HI has led to the formation of new stars in their tails \citep{kenney2014, poggianti2017, bellhouse2017, moretti2018}.
 
On the theoretical side, studies have also addressed the role that environment, and in particular RPS has on galaxy evolution \citep{bahe2013,bahe2015,steinhauser2016}, by monitoring their gas and H\,{\sc i} \citep{abadi1999, villaescusa2015,marasco2016,vandevoort2017}. Most recently, \citet{lotz2018} infers that RPS is the dominant mechanism that quenches infalling galaxies around their synthetic cluster sample. Cosmological hydrodynamical simulations \citep[e.g.][]{schaye2015,  pillepich2017} are an excellent tool for studying galaxy evolution as a function of environment, as they provide a full temporal and 6 dimensional phase-space description of all dark matter and baryonic material. However, in order to obtain realistic, highly resolved physics, cosmological simulations can be computationally expensive, often limiting their volume to periodic cubes with side length $\sim 100\text{Mpc}$. Therefore, though these simulations contain many groups ($\text{M}_{{\rm 200}} = 10^{13}-10^{14}\,h^{-1}\,\text{M}_{\odot}$), it is unlikely they will contain any massive clusters ($\text{M}_{{\rm 200}} >10^{15}\,h^{-1}\,\text{M}_{\odot}$\footnote{We define $\text{M}_{{\rm 200}}$ as the mass enclosed within a sphere of radius $\text{R}_{{\rm 200}}$ whose mean density is 200 times the critical density of the Universe.}). In light of this limitation, many recent theoretical efforts have been directed towards producing statistically significant sets of rich clusters using zoomed simulations; including but not limited to, {\sc Rhapsody-G} \citep{hahn2015}, {\sc Hydrangea} \citep{bahe2017}, {\sc C-Eagle} \citep{barnes2017}, and most recently, the \threehun\ project \citep{cui2018}. Simulations such as these have become important tools for studying galaxy evolution in high density environments, as we now know that environmental effects are not only of interest within the centre of rich galaxy clusters, but also in the faint outskirts, which is difficult to observe.

Recent studies have shown that galaxies are significantly altered from their initial state by group and filament environments, before even reaching the galaxy cluster; a phenomenon known as \textit{pre-processing} \citep{zabludoff1998,fujita2004,wetzel2013,cybulski2014, hou2014}. In fact, \citet{mcgee2009} used semi-analytic models to show that, on average, a $10^{14.5}\text{M}_{\odot}$ cluster at $z=0$ has accreted $40$ percent of its galaxies from group environments. Further to that, they suggested that galaxies being accreted into the cluster as part of a group are more massive than galaxies accreting from the field. However, pre-processing can again be observationally difficult to study due to contamination from backsplash galaxies \citep{gill2005,ludlow2009}. These are galaxies that have already entered the cluster core, undergone significant disruption and travelled back out to the cluster outskirts. \citet{gill2005} quantified this by using dark-matter-only simulations to show that $\sim50$ percent of galaxies residing between $\text{R}_{{\rm 200}}-2\,\text{R}_{{\rm 200}}$ of the main cluster dark matter halo are backsplash galaxies. 

However, both observational and theoretical studies have found success in studying environmental effects by using the \textit{phase-space} plane, which is constructed with the positions and velocities of each galaxy relative to it's cluster \citep{mahajan2011, jaffe2016,haines2015, oman2016, yoon2017}. The phase-space plane can be used to infer information about the assembly history of a particular cluster and the orbital histories of each galaxy around it, shedding more light on the link between environment and galaxy properties.

Using N-body simulations and full dark matter halo orbital histories, \citet{oman2013} showed that the time since infall of a halo into a cluster environment (defined as the time since first passage through $2.5\,{\rm R}_{{\rm 200}}$) is correlated with position on the phase-space plane. Alongside this, they also showed that different populations of haloes (e.g. backsplash, infalling and virialised) occupy distinct regions on the phase-space plane. By using all 6 dimensional quantities ($x,y,z,v_x,v_y,v_z$), simulations offer us a 6D view of the phase-space plane, but in observations we are limited to a projected view with 3 quantities (2 spatial and 1 orthogonal velocity). Comparisons between these views help to disentangle degeneracies that are inherently associated with projected data.

\citet{rhee2017} studied the stacked phase-space planes around 16 hydrodynamical zoom simulated clusters with mass range, $\sim 5\times10^{13}-9\times10^{14}\ h^{-1}\text{M}_{\odot}$. Using the orbital histories of infalling dark matter haloes, they found that the position on the projected phase-space plane is correlated with not only the time since infall of a particular halo, but also its tidal mass loss. They further concluded that this correlation evolves irrespective of cluster mass. However, from this study, it is not clear whether this statement holds for more-massive clusters, as the mass range and cluster counts are quite limited in their simulations. Two important questions remain: (i) is the phase-space position also correlated with the gas content of infalling haloes? If this is the case, this material can go on to form stars, which are directly observed and (ii) what is the primary mechanism that causes any type of mass (specifically gas mass) loss?

From an observational perspective, \citet{jaffe2015} attempted to answer these questions by using \ion{H}{I}-detected and non-detected galaxies around Abell 963, which is a $1.1\times10^{15}\ h^{-1}\text{M}_{\odot}$ galaxy cluster at $z=0.2$. They found that the likelihood of finding an H\,{\sc i}-detected galaxy in a particular region of the projected phase-space plane is inversely correlated with the intensity of their RPS model, implying that ram pressure plays a major role in galaxy H\,{\sc i} content. However, some questions remain. Firstly, the study finds a significant number of gas-poor, red galaxies in regions of low predicted RPS. Their spherically symmetric RPS model is based on an N-body simulation of a similar cluster and may not have accounted for these galaxies, as they lie further from the cluster where deviations away from the model become larger. A more comprehensive model could shed more light on this issue and also provide more detail nearer the cluster centre, as the accretion shock of clusters may not be spherically symmetric \citep{power2018}. Secondly, the strong correlation on the phase-space plane between H\,{\sc i} and the predicted RPS holds for this cluster, but it is not clear whether this holds for clusters with different masses and environments.

It is clear that the field is missing a theoretical study, analogous to the work done in \citet{jaffe2015}, which includes a statistically significant sample of galaxy clusters and a full treatment of cluster gas in each case. In this paper, we use \threehun\ project, a suite of 324 resimulated galaxy clusters containing a broad range of masses and environments, to investigate how the gas content of surrounding haloes correlates with phase-space position at $z=0$. These simulations provide a full description of the gas in not only the cluster, but also in filaments and group environments in the cluster outskirts. Using this set of simulations, we have built a ram pressure model that is individually calculated for each halo using its own local gas environment. We have investigated how the instantaneous ram-pressure is also correlated with phase-space position, and how this may be linked to halo gas content. Throughout this paper we only use information obtained from one snapshot, at $z=0$. Therefore, we may only know the instantaneous gas content of each halo and we refer to all haloes in our sample with low gas content as `gas-poor', as opposed to `quenched' or `stripped'. In order to explore whether haloes are indeed stripped, a full temporal analysis of each halo is needed, which will appear in a subsequent paper.

The paper is organised as follows. In \Sec{sec:methods} we provide details about how the data were obtained and descriptions of how the phase-space planes were constructed along with further calculations. We then present our results in \Sec{sec:results}. The first half of \Sec{sec:results} is devoted to the gas content of haloes in the phase-space plane and the second concentrates on the ram-pressure. \Sec{sec:conclusions} contains a summary of our main conclusions.

\section{Numerical Methods}\label{sec:methods}

\subsection{Hydrodynamical code}\label{sec:codes}
All resimulations used in this paper have been computed with the \gx\ code. \gx\ is a modified version of the non-public \gadgetthree\ code, which evolves dark matter with the \gadgetthree\ Tree-PM gravity solver \citep[an updated version of the \gadgettwo\ code;][]{springel2005b} and uses a modern Smoothed Particle Hydrodynamics (SPH) scheme to provide a full treatment of the gas particles. This SPH scheme includes: an artificial conduction term that largely improves the SPH capability of following gas-dynamical instabilities and mixing processes, a higher-order Wendland C4 kernel \citep{dehnen2012} to better describe discontinuities and reduce clumpiness instabilities, and a time-dependent artificial viscosity term to minimise viscosity away from shock regions. The performance of this SPH algorithm is presented in \citet{beck2016} and \citet{sembolini2016}, where the latter work specifically focuses on the performance in galaxy cluster simulations. 

As the spatial and temporal range of more complex baryonic processes is too large for current computational facilities and codes to solve self-consistently, hydrodynamical codes compute these with analytic prescriptions that contain tuneable free parameters, also known as subgrid physics. The subgrid physics schemes contained in \gx\ are as follows:
\begin{itemize}
\item \textit{Gas cooling and heating}: Gas cooling is computed for an optically thin gas and takes into account the contribution of metals, using the method in \citet{wiersma2009}, while a uniform UV background is included by following the procedure of \citet{haardt1996}.
\item \textit{Star Formation}: Star formation is carried out as in \citet{tornatore2007}, and follows the star formation algorithm presented in \citet[][henceforth, SH03]{springel2003}, where gas particles above a given density threshold are treated as multi-phase. The SH03  model describes a self-regulated inter-stellar medium and provides a star formation rate that depends upon only the gas density.
\item \textit{Stellar Population Properties \& Chemistry}: Each star particle is considered to be a single stellar population (SSP). The evolution of each SSP is followed according to \citet{chabrier2003}. Metals are accounted for in the Type Ia and Type II supernovae and AGB
phases, and the code follows 16 chemical species. Star particles are stochastically spawned from parent gas particles as in SH03, and get their chemical composition from their parent gas. 
\item \textit{Stellar Feedback}: Supernova feedback is modelled as kinetic and the prescription described in SH03 is followed (i.e. energy-driven scheme with a fixed wind velocity of 350$\,\text{km}\,\text{s}^{-1}$, wind particles decoupled from surrounding gas for a period of 30 Myr or until ambient gas density drops below 0.5 times the multiphase density threshold).
\item \textit{Super Massive Black Hole (SMBH) Growth \& AGN Feedback}: AGN feedback follows \citet{steinborn2015}, where SMBHs grow via the Eddington-limited Bondi--Hoyle-like gas accretion, with the model distinguishing between a cold and hot component. 
\end{itemize}

\gx, equipped with these subgrid schemes, has been extensively tested by the Trieste numerical group \citep{rasia2015,planelles2017,biffi2017} and in the \textit{nIFTy cluster comparison} project \citep{sembolini2016b,elahi2015,cui2016b,arthur2017}. These latter studies have shown that \gx\ was able to produce a realistic and stable $\text{M}_{\text{200}}=1.1\ \text{x}\ 10^{15}\ h^{-1}\text{M}_{\odot}$ cluster at $z=0$, using a similar particle mass resolution as used in \threehun\ project. 

\begin{figure*}
    \centering
    \includegraphics[scale=0.592]{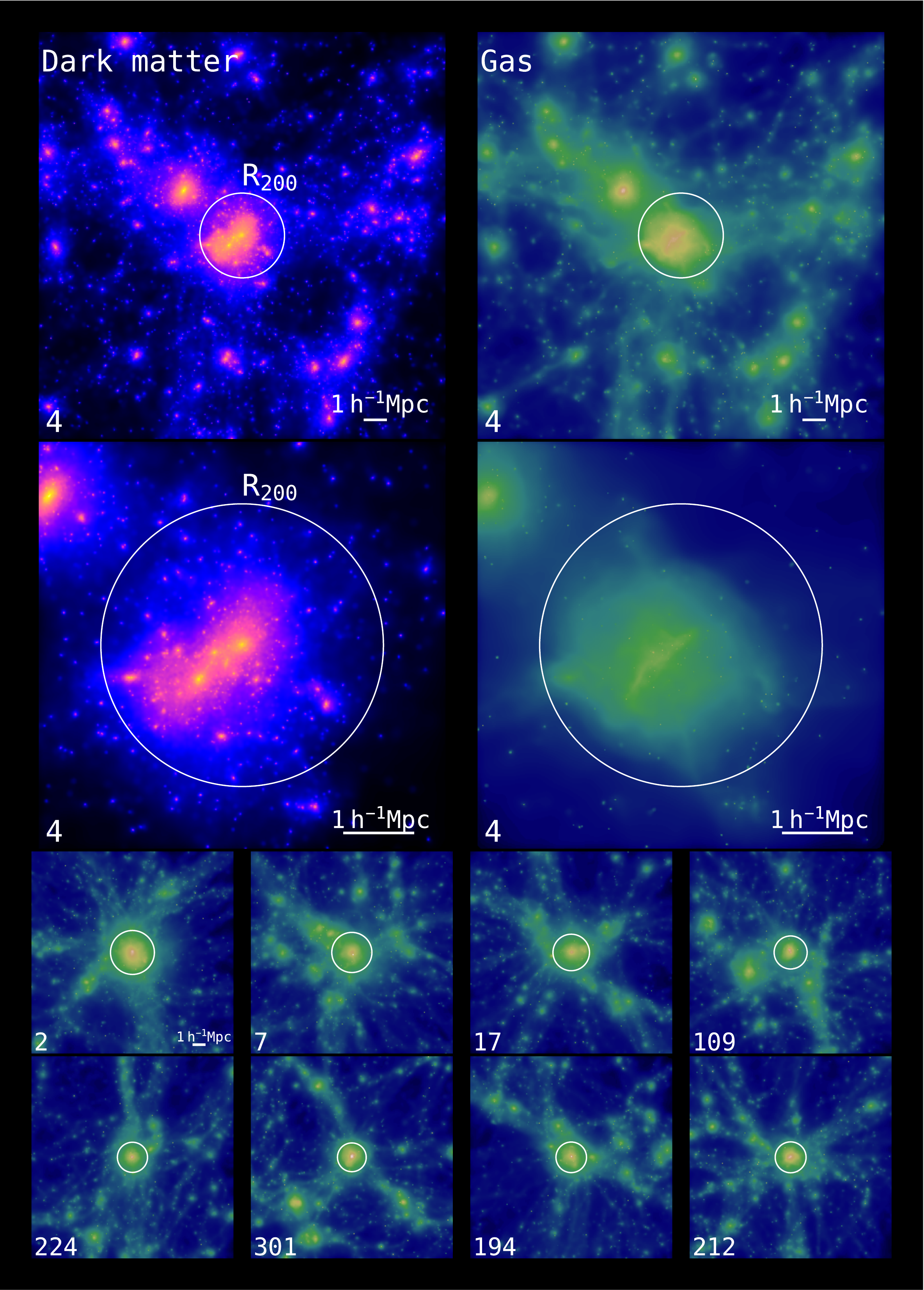}
    \caption{The figure shows nine example clusters from the \threehun\ project at $z=0$. The top four panels are common to the same cluster, whilst the bottom eight panels each show a different cluster. The number in the bottom-left of each panel corresponds to the cluster number in the \threehun\ database. For cluster 4, the left and right panels show the projected dark matter and gas distributions respectively. For all other clusters, only the projected gas distribution is shown. Two perspectives are shown for cluster 4. The top panels show a $ 20\ h^{-1}\ \text{Mpc}\times20\ h^{-1}\ \text{Mpc}$ view of the simulation (this projection is also used for the clusters in the bottom eight panels), whilst the bottom panels shows a $6\ h^{-1}\ \text{Mpc}\times6\ h^{-1}\ \text{Mpc}$. The white circle in each panel respresents where ${\rm R}_{{\rm 200}}$ is for each cluster halo. The figure demonstrates the suitability of the data for this study, as it contains a statistically significant sample of massive clusters embedded in a wide range of environments, a full treatment of the gas and a large resimulation volume that includes many filamentary and group structures around each cluster.}
    \label{fig:mosaic}
\end{figure*}

\subsection{Galaxy cluster resimulations}\label{sec:data}
In this paper we use 324 resimulated galaxy clusters from \threehun\ project \citep{cui2018}. These are zoom simulations based on massive haloes that were initially extracted from the dark-matter-only MDPL2, MultiDark simulation \citep{klypin2016}. MDPL2 was simulated with \textit{Planck} cosmology ($\Omega_{\rm M}\,=\,0.307,\,\Omega_{\rm B}\,=\,0.048,\,\Omega_{\Lambda}\,=\,0.693,\,h\,=\,0.678,\,\sigma_{8}\,=\,0.823,\,n_{\rm s}\,=\,0.96$) and consists of a periodic cube of comoving side-length $1\ h^{-1}\ \text{Gpc}$, which contains $3840^3$ dark matter particles, each of mass $1.5\ \text{x}\ 10^9\ h^{-1}\text{M}_{\odot}$ \citep{planck2016}. 

In short, the most massive 324 haloes from MDPL2 at $z=0$ form the basis of the \threehun\ dataset. These haloes, and the regions around them, were taken back to their initial conditions and resimulated down to $z=0$ with \gx. Further details about how this dataset was obtained can be found in \citet{cui2018}, including further selection criteria and resimulation pipelines.

The subsequent resimulations contain a high-resolution volume, specifically a sphere of radius $15\ h^{-1}\ \text{Mpc}$ centred on each cluster and contains dark matter and gas particles with mass $12.7\ \text{x}\ 10^8\ h^{-1}\text{M}_{\odot}$ and $2.36\ \text{x}\ 10^8\ h^{-1}\text{M}_{\odot}$ respectively. On average, this region extends beyond $5\,\text{R}_{\text{200}}$ of the galaxy cluster dark matter halo. In many instances, the volumes include filaments and groups feeding the cluster from the outskirts. For each galaxy cluster, 129 snapshots from $z=16.98$ to $z=0$ were saved. At $z = 0$, the 324 galaxy clusters have a mass range from $\text{M}_{\text{200}}=6.08\ \text{x}\ 10^{14}\ h^{-1}\text{M}_{\odot}$ to $\text{M}_{\text{200}}=2.62\ \text{x}\ 10^{15}\ h^{-1}\text{M}_{\odot}$.

\Fig{fig:mosaic} shows an example of several galaxy clusters from \threehun\ project at $z=0$. The figure contains the projected density views of 9 different clusters in both dark matter and gas. We highlight here the wealth and diversity of information contained in \threehun\ project. Apart from the fact that it contains many massive clusters, we note that many of the clusters inhabit a broad range of environments; some appear more relaxed and isolated, whereas others appear to be undergoing major mergers whilst being fed by many groups and filaments in the infall region. Visualisations and movies of all 324 resimulations at different epochs are located at \href{http://music.ft.uam.es/videos/music-planck}{\url{http://music.ft.uam.es/videos/music-planck}}.

\subsection{Halo catalogues}
All haloes and subhaloes were found and analysed with the \ahf\ \footnote{\url{http://popia.ft.uam.es/AHF}} halo finder \citep{knollmann2009}, which self-consistently includes both gas and stars in the halo finding process  \citep[for further information about the perfomance of \ahf, see:][]{knebe2011}. In this paper we utilise the following definitions:

\begin{itemize}
\item $R_{200}$: The radius which is equal to $\text{R}_{\text{200},\,\text{crit}}$ of the cluster halo. ($\text{R}_{\text{200},\,\text{crit}}$ is the radius of a sphere whose mean density is 200 times the critical density of the Universe.)
\item $R_{200,\,h}$: When used this refers to $\text{R}_{\text{200},\,\text{crit}}$ of any halo or subhalo in question (excluding the cluster halo).
\item \textit{Halo}: An object comprised of a collection of dark matter and baryonic particles, as classified by \ahf. In our definition, a halo \textit{does not} reside within $\text{R}_{\text{200},\,\text{h}}$ of another halo or $\text{R}_{\text{200}}$ of the cluster halo.
\item \textit{Subhalo}: A halo that \textit{does} reside within $\text{R}_{\text{200},\,\text{h}}$ of another halo, be it another subhalo or halo, or $\text{R}_{\text{200}}$ of the cluster halo.	
\item \textit{Cluster halo}: The most massive halo in each resimulation volume at $z=0$. Each resimulation volume is centred on this object. 
\end{itemize}

We discard all haloes and subhaloes from our analysis below $\text{M}_{\text{200}}=10^{11}\ h^{-1}\text{M}_{\odot}$, as these objects contain $\lesssim\,100$ particles; and any halo at a distance of $>5\,\text{R}_{\text{200}}$ from their corresponding cluster. 169287 haloes and subhaloes satisfy these criteria and make up our final sample.

In our analysis we use the gas fractions of each halo and subhalo, $\text{f}_{\text{g}}$, which is simply the mass of all the gas particles within $\text{R}_{\text{200}}$ of an object over the total mass of the same object.

\subsection{Phase space}\label{sec:phase-space}
All phase-space coordinates in this paper have been calculated in a similar fashion to \citet{oman2013}. The planes are constructed in 6D and a Line-Of-Sight (LOS) projection, where the latter is designed to mimic observational studies. In order to obtain the phase-space coordinate of an object at $z=0$, one must know; the distance of each object from the galaxy cluster centre (defined as the peak in the adaptively smoothed density field computed by \ahf), $\text{R}$, and the relative velocity between the object and the galaxy cluster halo, $\text{v}$. Distances are normalised by the virial radius of the galaxy cluster halo, $\text{R}_{\text{200}}$, and relative velocities by the cluster velocity dispersion, $\sigma$. 

\begin{figure*}
    \centering
    \includegraphics[width=\textwidth]{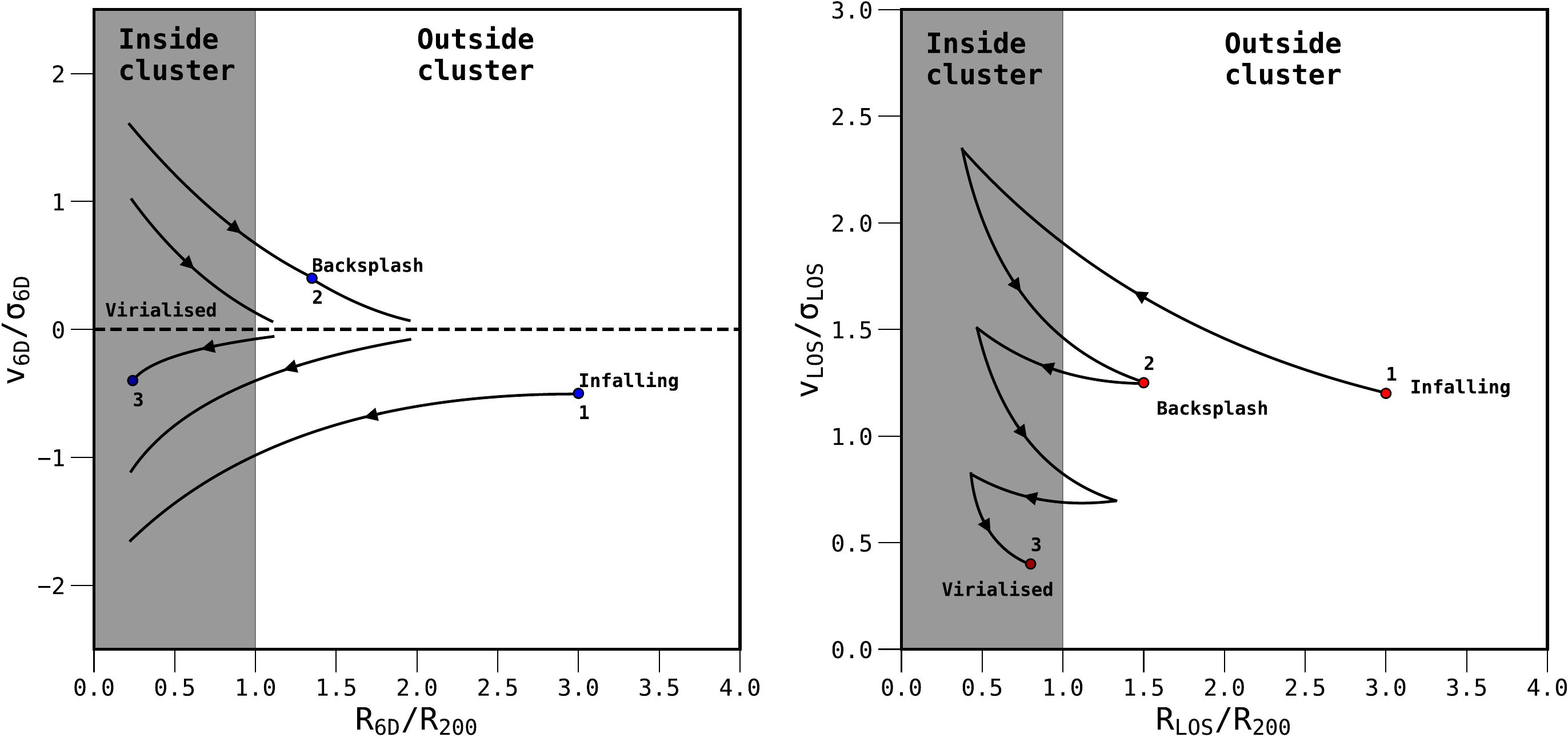}
    \caption{A schematic showing the typical evolution of a halo or subhalo over a 6D (left-hand panel) and LOS (right-hand panel) phase-space plane with time. The trajectories shown in both planes are not supposed to represent the same object trajectory in different perspectives. The diagram shows how different regions of the phase-space plane contain different populations of haloes and subhaloes (e.g. backsplash, infalling and virialised), and the differences between a 6D ('true') view and a LOS ('observed') projection. See text for more details. Note that we show the time evolution of a typical object here for conceptual purposes; however, for the remainder of this study, we strictly limit our analysis to $z=0$.}
    \label{fig:schematic}
\end{figure*}

\Fig{fig:schematic} shows a schematic of how a typical halo evolves over a 6D and LOS phase-space plane with time. The typical (sub)halo trajectory in both cases starts with an infalling object that moves closer to the cluster whilst increasing in speed. As an object approaches the pericentre of its orbit, the velocity sign is flipped in the 6D projection and the object can move away from the cluster to form the backsplash population that is distinct from the first infalling population. However, in the LOS perspective, objects that move away from the cluster after their pericentre passage form a backsplash population that is almost indistinct from the first infalling population. Lastly, on both planes each object becomes virialised at their end of their trajectory.

\subsubsection{6D}

The 6D phase-space coordinate of a halo or subhalo is found using all 6 phase-space dimensions of the object ($x_h,\ y_h,\ z_h,\ v_{x,h},\ v_{y,h},\ v_{z,h}$) with respect to the same dimensions of the cluster halo ($x_c,\ y_c,\ z_c,\ v_{x,c},\ v_{y,c},\ v_{z,c}$). Specifically, the 6D velocity of an object, ${\rm v}_{\rm 6D}$, is found with,
\begin{equation}
\label{v_6d}
{\rm v}_{\rm 6D} = {\rm sgn}(\bm{{\rm v}}\cdot\bm{{\rm r}})\,|\bm{{\rm v}}|,
\end{equation}
where $\bm{{\rm r}}$ and $\bm{{\rm v}}$ are the position and velocity vectors between the cluster and the object respectively. The sign of ${\rm v}_{\rm 6D}$ allows us to disentangle which objects are infalling into or outgoing from the cluster at $z=0$. $\sigma_{\rm 6D}$ is found by taking the root mean square of the ${\rm v}_{\rm 6D}$ distribution of subhaloes in the cluster halo that have been defined by \ahf.

\subsubsection{LOS}

We have also calculated the LOS phase-space coordinate of each object in order to study how projection effects alter our results. Here, we arbitrarily use one LOS projection down one axis of our simulation box, though we note that our results are not sensitive to which one is chosen. From this projection, two spatial and the orthogonal velocity component are used (e.g. $x,\ y,\ v_z$) to calculate the distance from the cluster,
\begin{equation}
\label{r_los}
{\rm R}_{\rm LOS} = \sqrt{(x_h-x_c)^2+(y_h-y_c)^2},
\end{equation}
and the projected velocity with a correction for the Hubble flow,
\begin{equation}
\label{v_los}
{\rm v}_{\rm LOS} =|(v_{z,h}-v_{z,c})-H(z_h-z_c)|.
\end{equation}
$\sigma_{\rm LOS}$ is computed as the root mean square of the ${\rm v}_{\rm LOS}$ distribution of subhaloes in projection (${\rm R}_{\rm LOS}<{\rm R}_{\rm 200}$). We find that if we consider objects within different radii in our velocity dispersion estimate, such as objects within ${\rm R}_{\rm LOS}<2\,{\rm R}_{\rm 200}$, our velocity dispersions are biased low because objects in the outskirts are moving slower. However, this bias is only $\lesssim 3$ percent of our original values. 

As previously stated, we only use objects within $5\,\text{R}_{\text{200}}$ of the cluster halo in each resimulation and do not have a full lightcone through a cosmological size simulation box. However, our radial selections are somewhat similar to the LOS spread in the z-direction one might receive if they were to make a typical photometric selection of the cluster members. 

Once the phase-space information has been computed for each cluster, the planes are binned onto a $50\times50$ grid, and can be stacked. We stack our planes in the ranges, $0\,<\,{\rm R}/{\rm{R_{{\rm 200}}}}\,<\,4$ and $-2.5\,<\,{\rm v}/{\rm \sigma}\,<\,2.5$. The bin sizes are therefore, $\delta({\rm R}/{\rm{R_{{\rm 200}}}})=0.08$ and $\delta({\rm v}/{\rm \sigma})=0.1$.

\subsection{Ram pressure}\label{sec:ram pressure}
For the purposes of this paper, we calculate the instantaneous ram pressure acting upon a (sub)halo as in \citet{cunnama2013}; by using,
\begin{equation}
\label{pram}
\text{P}_{\text{ram}} = \rho_{\text{gas}}v_{\text{ram}}^2.
\end{equation}
Here, $\rho_{\text{gas}}$ is the mass density of all gas particles in a shell centred on the relevant (sub)halo. $v_{\text{ram}}$ is the magnitude of the difference between the (sub)halo velocity and the median bulk flow for the same gas particles. The shell around each (sub)halo arbitrarily extends between  $1 - 2\ \text{R}_{\text{200},\,\text{h}}$. However, we note that changing these radii does not alter the instantaneous ram pressure significantly. In fact, there is only a $\lesssim 6$ percent deviation in the median ram pressure for all (sub)haloes in our sample between our shell boundaries ($1-2\ \text{R}_{\text{200},\,\text{h}}$) and other sample boundaries  ($1.5-2.5\ \&\ 2-3\ \text{R}_{\text{200},\,\text{h}}$).

\section{Results \& Discussion}\label{sec:results}

We split our results into two parts. Firstly, we present the gas content of both haloes and subhaloes, quantifying the halo or subhalo gas fraction as a function of phase-space coordinate. We stack the data from all 324 of our re-simulations to obtain ensemble averages. We repeat the analysis in the second part, concentrating instead on the instantaneous ram pressure felt by each halo (or subhalo).

\begin{figure*}
    \begin{subfigure}{\textwidth}
    	\centering
        \includegraphics[scale=0.41]{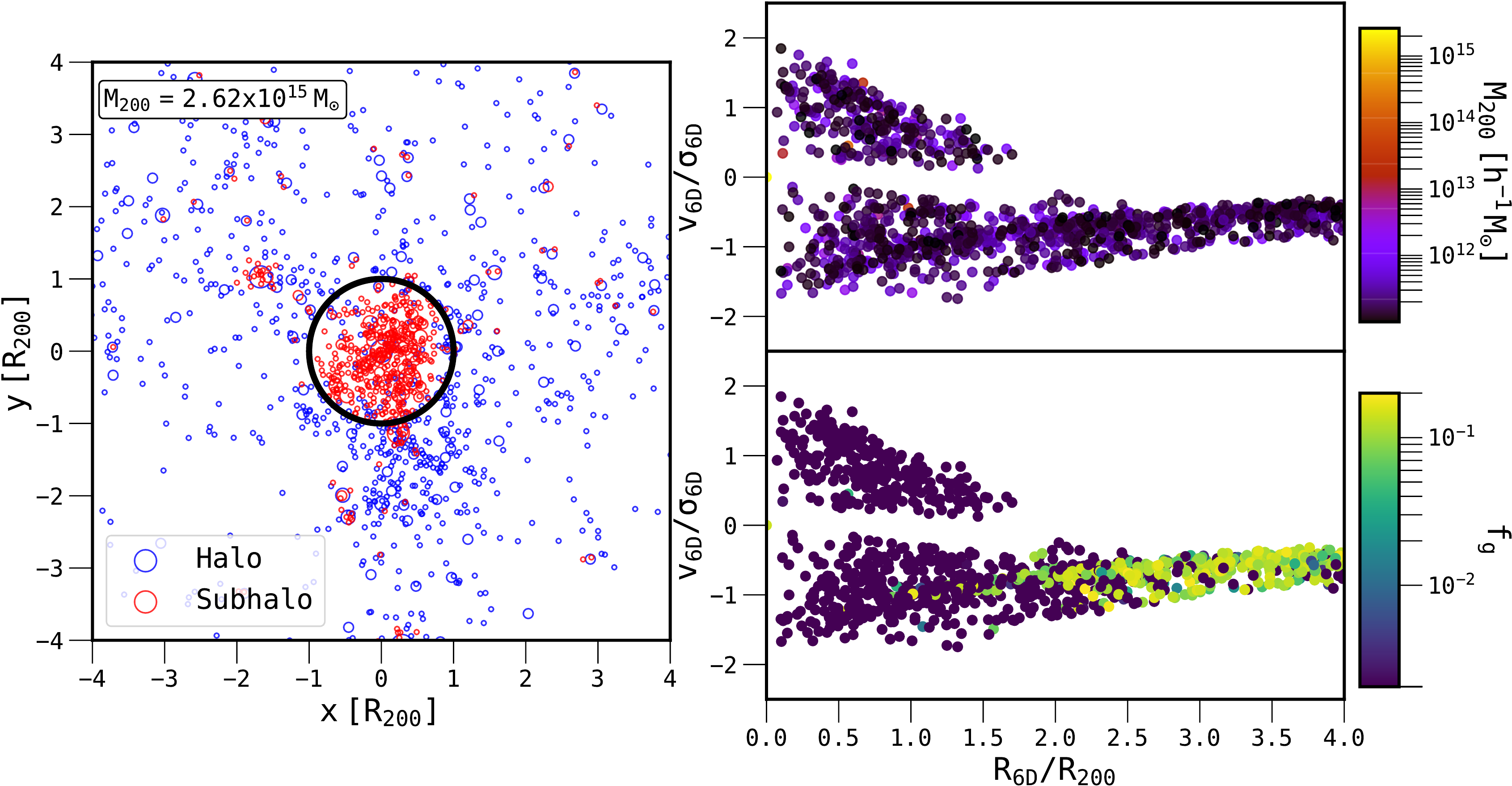}
        \label{fig:clus_relax}
    \end{subfigure}

    \begin{subfigure}{\textwidth}
    \centering
        \includegraphics[scale=0.41]{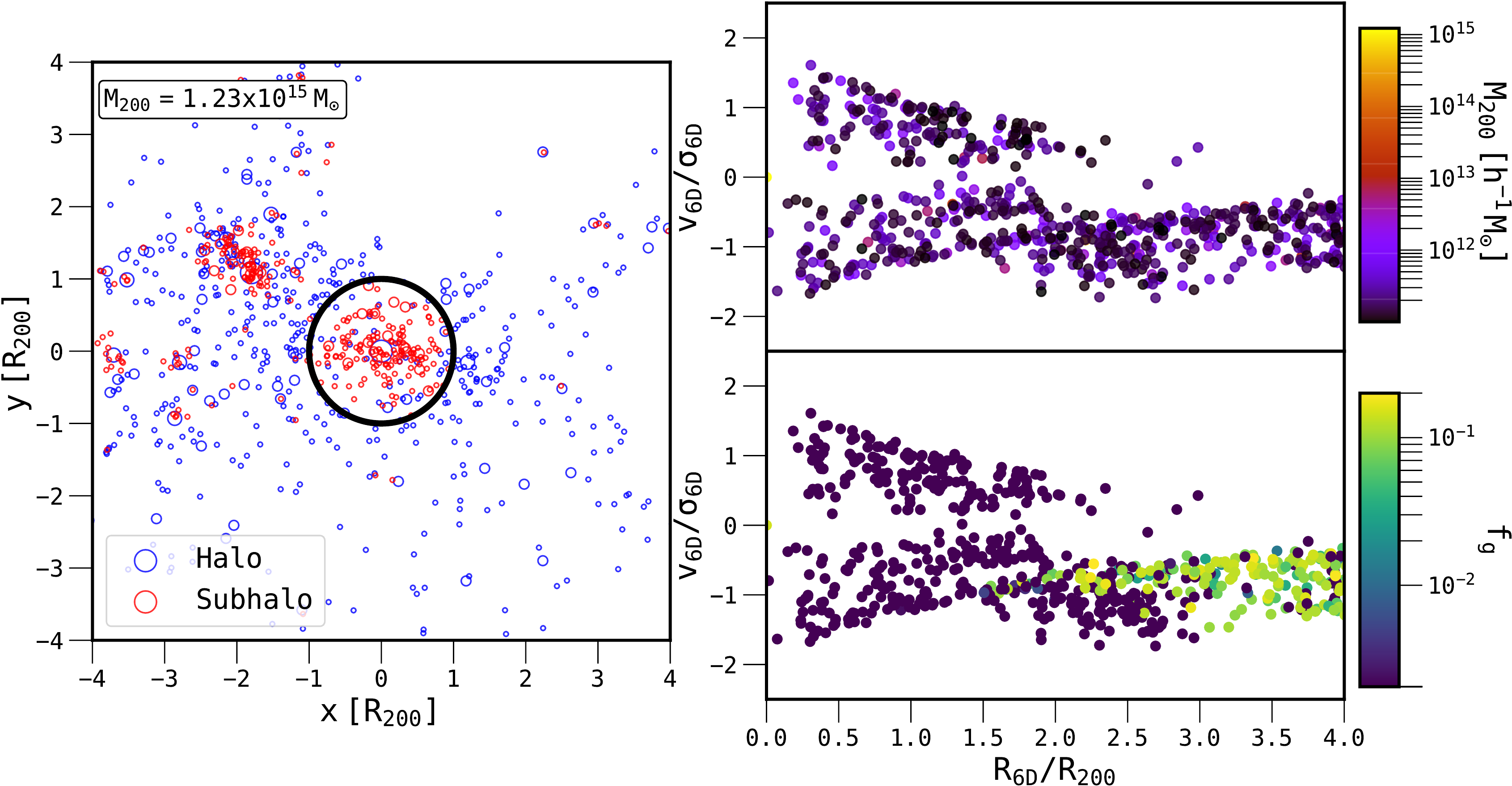}
        \label{fig:clus_merge}
    \end{subfigure}

    \begin{subfigure}{\textwidth}
    \centering
        \includegraphics[scale=0.41]{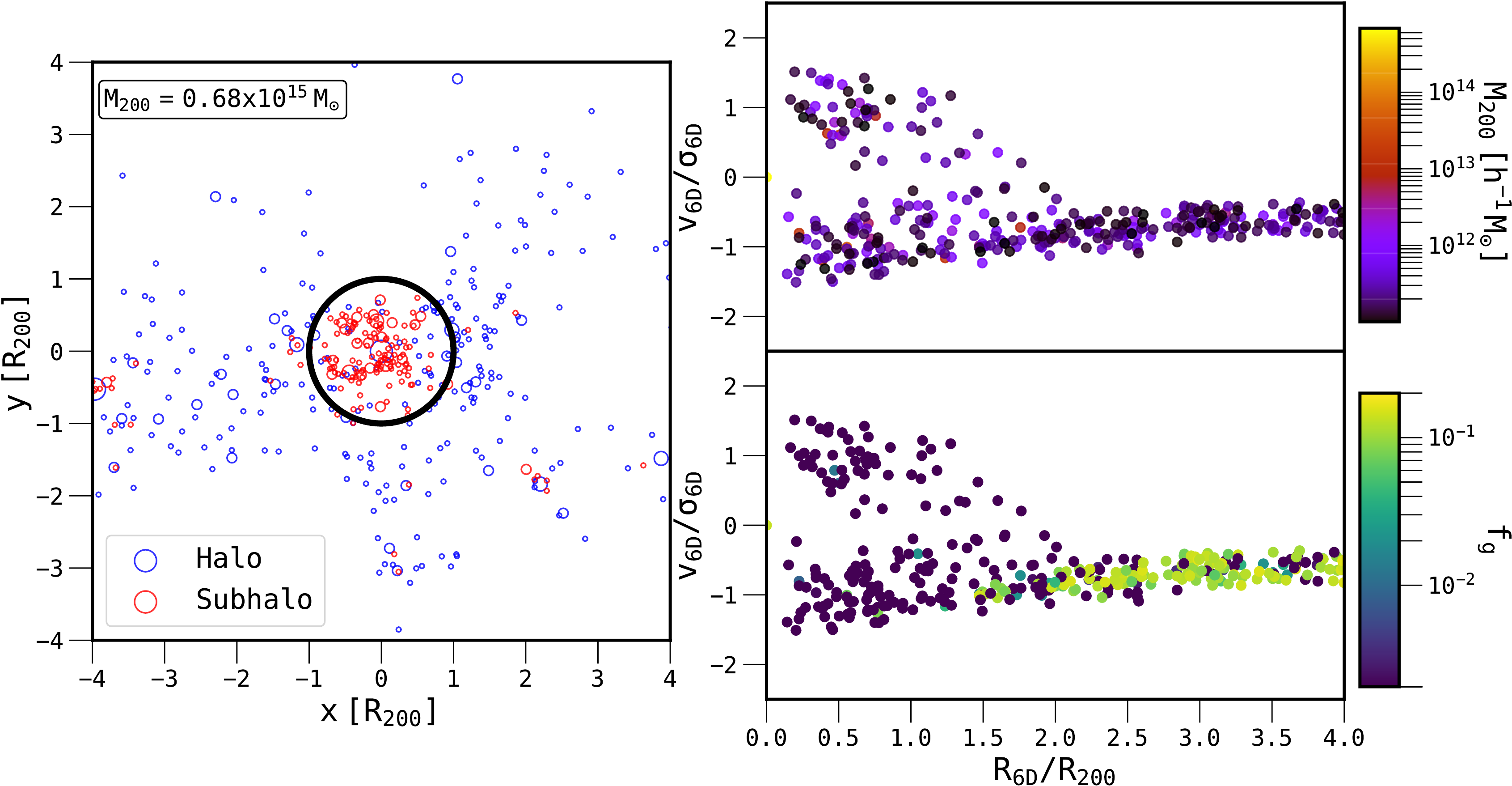}
        \label{fig:clus_lowmass}
    \end{subfigure}
    \caption{Three sample cluster resimulations from the \threehun\ project are shown. For each resimulation, the left panel shows the projected halo and subhalo spatial distributions in blue and red respectively. The black circle indicates where $\text{R}_{\text{200}}$ of the cluster in the centre is. In the right panels, the 6D phase-space information for all of these haloes and subhaloes can be found. The top and bottom phase-space panel for each region are colour-coded by $\text{M}_{\text{200}}$ and gas fraction respectively. \threehun\ project enables access to this information in a range of cluster environments, and it is apparent in these examples that the halo and subhalo gas fraction may be correlated with phase-space location.}
\label{fig:sing_clus} 
\end{figure*}

\subsection{Gas content}\label{sec:gas_content}

\subsubsection{Case study: the phase-space plane for individual clusters}
To illustrate the power of our dataset and the information it contains, in \Fig{fig:sing_clus} we display three different clusters from \threehun\ project, at $z=0$. The left-hand panels show the projected halo and subhalo distributions, whilst the two right-hand panels display the 6D phase-space planes for all objects seen in projection. We have selected these three clusters as they portray three different large-scale cluster environments with varying cluster halo masses. 

The top panel depicts one of the most massive clusters in our sample. It is relaxed, with no on-going major merger or secondary large haloes nearby. The outskirts contain little substructure at $z=0$, and most of the infalling material at this snapshot is isolated. By contrast, the middle panel shows a cluster in quite a different environment. Though it has a similar mass to the cluster above, it appears less relaxed: one can see the cluster halo is in close proximity to another large object that contains significant amounts of substructure in the top-left quadrant of the panel. This could indicate that the cluster is about to undergo a merger, or that it already has and the material has punched through to the other-side. Regardless, it is clear that these unrelaxed regions in the outskirts are likely places to look for evidence suggesting pre-processing. Finally, the bottom panel shows a cluster that is somewhat less massive and is being fed by significantly less material than the other two examples. From this projected view, it appears as though a filament on the left-hand side is connected to the cluster halo and subsequently feeding it. 

On the right-hand side of \Fig{fig:sing_clus} we provide the phase-space distribution of the haloes and subhaloes colour-coded by either mass or gas fraction as indicated. By considering the phase-space planes of these clusters, it is interesting to note how gas-poor ($\text{f}_{\text{g}}<10^{-2}$) haloes appear to be nearer the cluster centre. At around $\sim 1.5-2\ \text{R}_{\text{200}}$, many of the haloes seem to have lost nearly all of their gas. In all three cases, $\gtrsim 75$ percent of haloes are gas-poor within $\sim\,1.5-2\ \text{R}_{\text{200}}$. This transition is specifically seen for objects with ${\rm v}_{\rm 6D}/\sigma_{\rm 6D}<0$, and at  $\sim\,1.5-2\ \text{R}_{\text{200}}$: many of these objects will be on their first infall into the cluster \citep{oman2013}. Therefore, this suggests objects are becoming gas-poor as they first infall, which is consistent with the findings of \citep{lotz2018}.

However, these phase-space planes also highlight differences between these clusters as well. Outside $\text{R}_{\text{200}}$, there are also gas-poor haloes in the top two planes, but in this region $\sim 19$ percent are gas-poor in the top panel, whilst $\sim 30$ percent are in the middle panel. Objects in the outskirts are also more dynamic in the middle panel compared to the top panel, and it is safe to assume that the highly substructered outskirts around the cluster in the middle panel are driving these differences. In particular, the widening of object phase-space velocities between $\sim\,2-2.5\ \text{R}_{\text{200}}$ for the middle cluster is due to the group environments seen in the top-left quadrant in the left-hand panel. It is important to note that there is a wide cluster-to-cluster variation visible within these three examples and that only by stacking many objects can we hope to robustly extract general trends.

\begin{figure}
    \centering
    \includegraphics[width=\columnwidth]{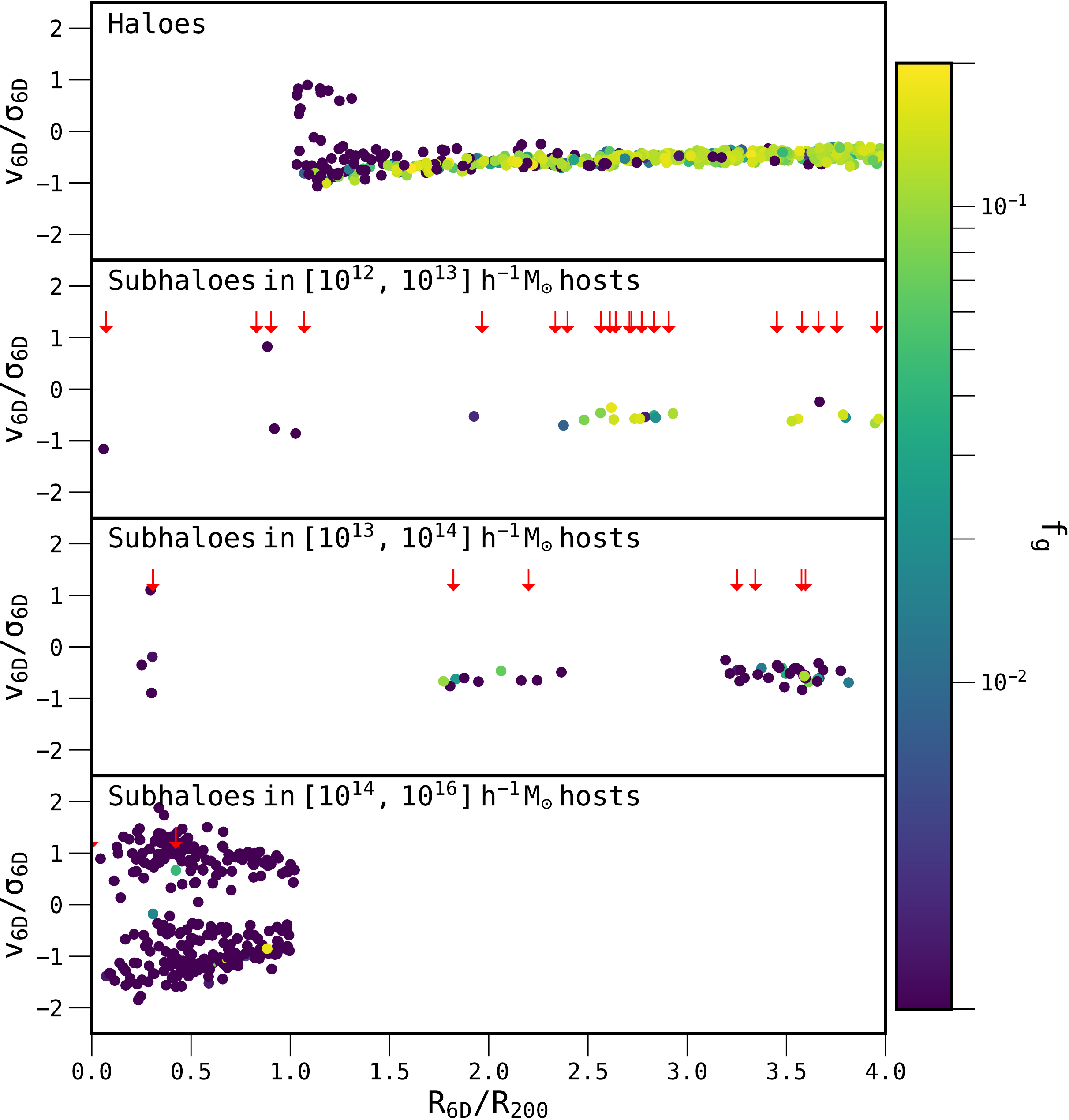}
    \caption{The phase-space plane for all haloes and subhaloes residing within 4\ $\text{R}_{\text{200}}$ of cluster 0005 (an example case) in the \threehun\ project. The top panel shows the phase-space information for the haloes in and around the cluster, whilst the bottom three panels show subhaloes; each panel is binned by the host mass that the subhaloes reside in, as indicated at the top-left of each panel. Each halo and subhalo are colour-coded by gas fraction. The red arrows in the bottom three panels mark the positions of the host haloes that the subhaloes reside in. In this example subhaloes are more gas-poor and their gas fractions decrease as the host halo's mass increases.}
    \label{fig:sing_clus_gfrac_substat}
\end{figure}

We can also slice the data for a single resimulation, splitting the sample into haloes and subhaloes, further subdividing the subhaloes either directly in mass or via the mass of their host halo. This will allow us to examine the gas content of main haloes in the phase-space plane as well as addressing whether or not subhaloes are principally influenced by their position relative to the main halo or their presence within a larger structure. We illustrate these effects for a single re-simulation in \Fig{fig:sing_clus_gfrac_substat} (in this instance cluster 0005, an example case). Here, we show the 6D phase-space plane, of \Fig{fig:sing_clus}, only for the gas fraction. As in the bottom-right panels of that figure this phase-space plane is also colour-coded by gas fraction. However, in this figure we split the phase-space plane up into four panels based on whether they are haloes or subhaloes. The subhaloes are also binned further, depending on the mass of their host halo. 

\begin{figure*}
    \centering
    \includegraphics[width=0.8\textwidth]{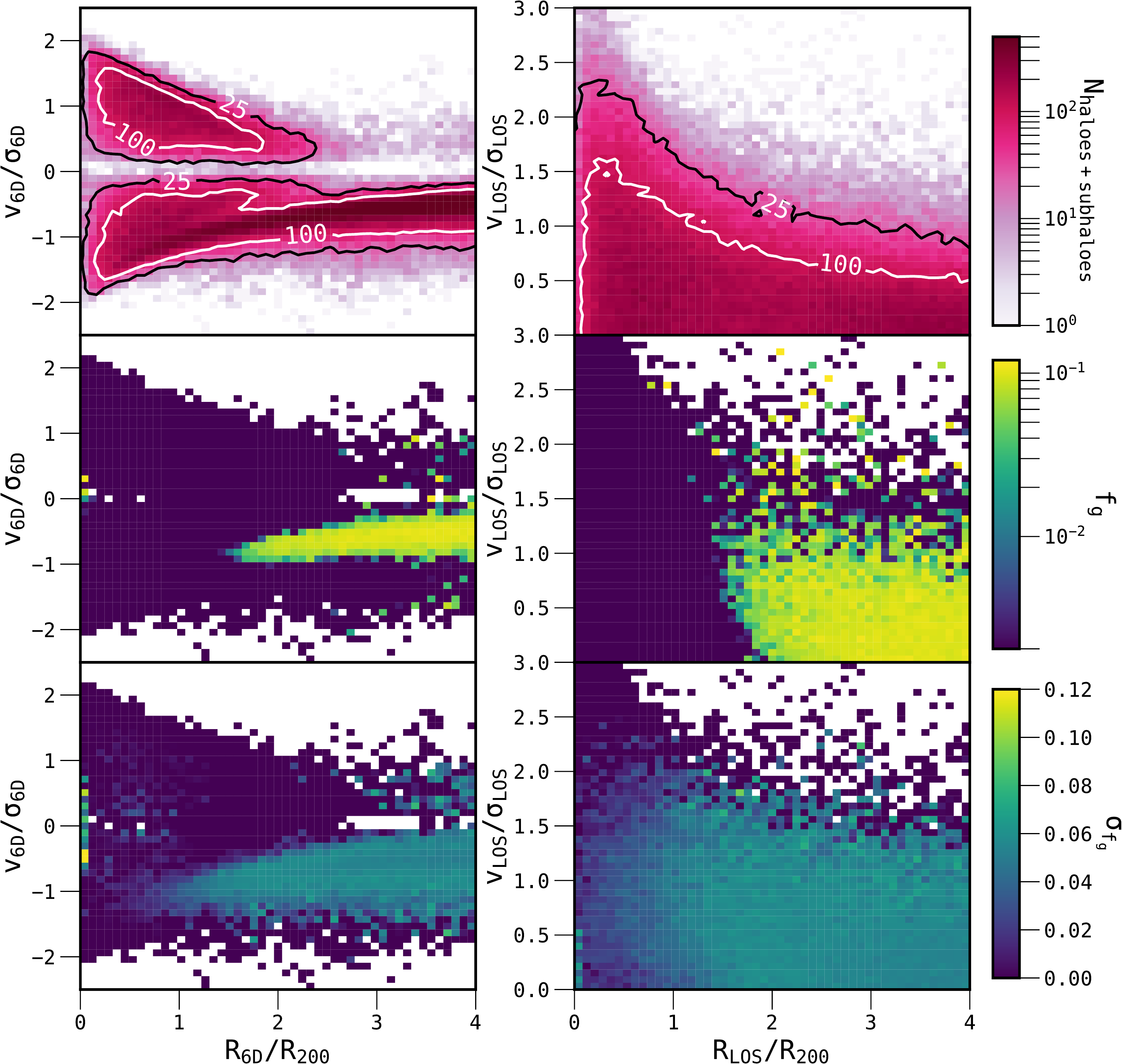}
    \caption{The stacked phase-space plane for all 324 clusters in the \threehun\ project. Both haloes and subhaloes are included in all panels. The left and right columns are in 6D and LOS perspectives respectively. The planes are stacked on a $50\times50$ grid where the top row shows the number of haloes and subhaloes in each bin, the middle row indicates the median gas fraction in each bin and the bottom row illustrates the gas fraction standard deviation in each bin. Contours are included in the top row, indicating the number of haloes and subhaloes encapsulated in each. The 6D perspective shows two distinct infalling and outgoing populations of objects, which is not seen in the LOS perspective. Also, from the the 6D perspective, first infallers are substantially more gas-poor and appear to lose their gas at $\sim\,1.5\ \text{R}_{{\rm 200}}$, whereas the cut-off in gas rich objects in the LOS perspective occurs closer to $\sim\,2\ \text{R}_{{\rm 200}}$.}
    \label{fig:all_stack}
\end{figure*}


\begin{figure*}
    \centering
    \includegraphics[width=\textwidth]{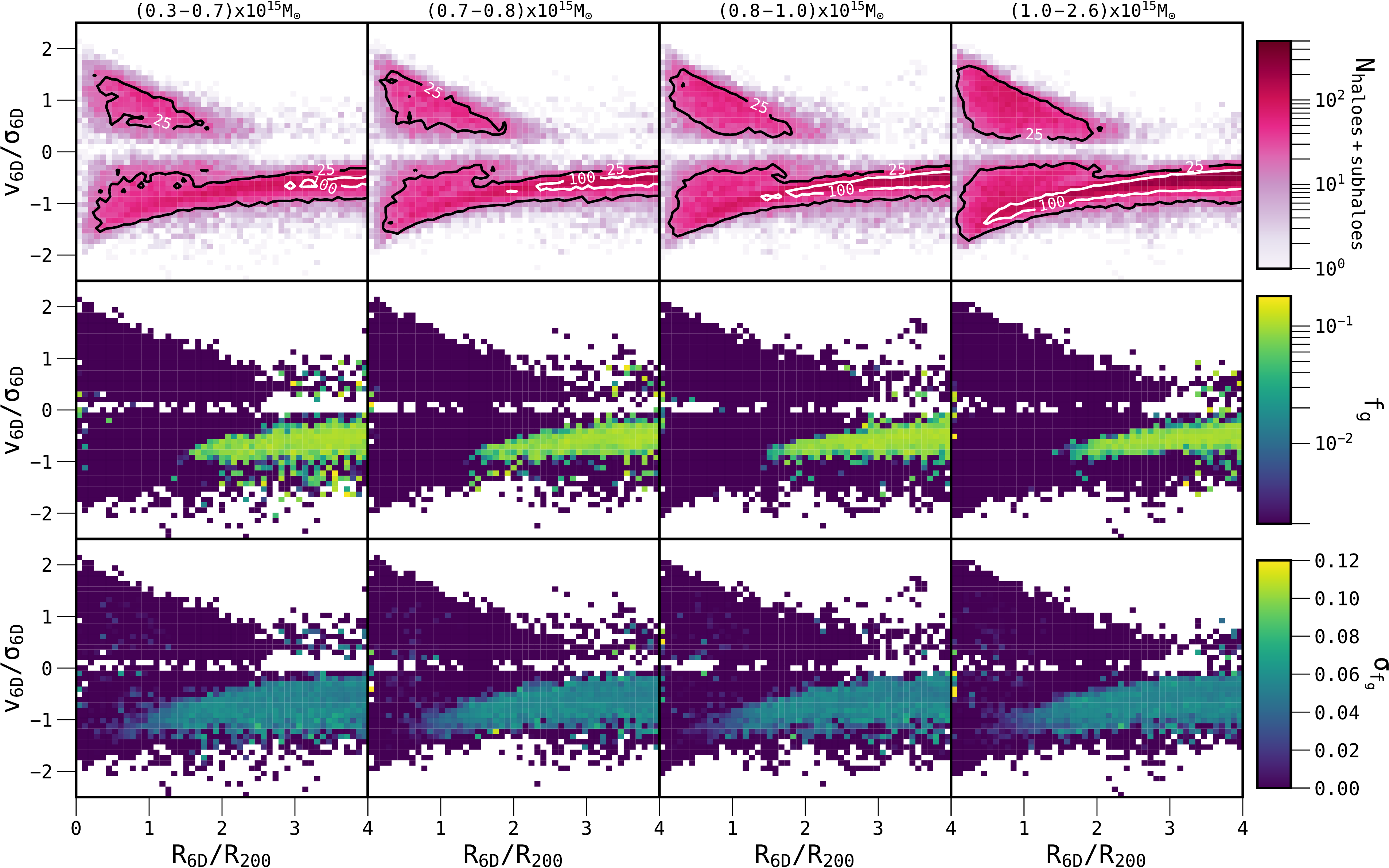}
    \caption{As in \Fig{fig:all_stack}, but only shown for the 6D perspective and also binned by cluster mass. Each column represents one cluster mass bin, where the mass range is indicated above the top panel. The bin edges were chosen so that each bin contains 81 clusters. It is clear that as the cluster mass increases, the number of haloes and subhaloes on the plane also increase. The shape of the plane is not drastically different between bins. In all bins there is a sharp cut-off in the gas richness of infalling objects at $\sim\,1.5\ \text{R}_{{\rm 200}}$, which seems independent of cluster mass.}
    \label{fig:clus_mass_bins}
\end{figure*}

\begin{figure*}
    \centering
    \includegraphics[width=0.7\textwidth]{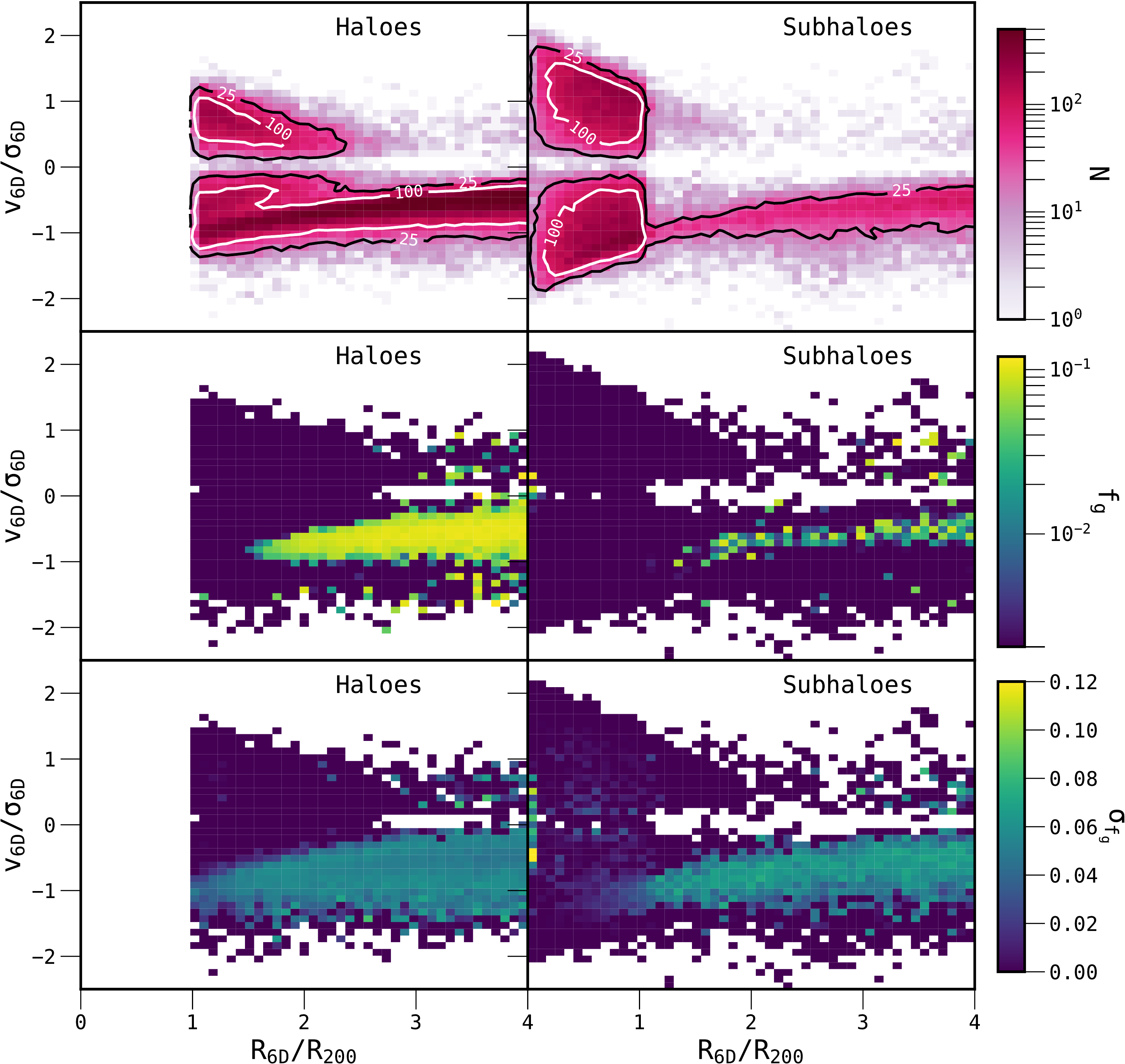}
    \caption{The stacked phase-space planes for all 324 clusters in the \threehun\ project, split by subhalo status. The left- and right-hand columns show only haloes and subhaloes respectively in the 6D perspective. The planes are constructed like in \Fig{fig:all_stack}. $\sim\,10$ percent of all infalling objects are in the form of subhaloes, which are substantially more gas-poor than haloes in the same phase-space location.}
    \label{fig:all_stack_substat}
\end{figure*}


From this example cluster, we see that, on average, subhaloes are less gas-rich than the halo population. This is especially true for subhaloes with increasing host mass; where for instance, $\sim\,77$ percent of subhaloes with group-like host mass $10^{13}\ h^{-1}\text{M}_{\odot}-10^{14}\ h^{-1}\text{M}_{\odot}$ are gas-poor, even well beyond ${\rm R}_{\rm 200}$. However, for this cluster case it is unclear whether gas richness is entirely dictated by an object's subhalo status, as $\sim 27$ percent of haloes in the top panel are gas-poor. Ultimately though, in this example subhaloes are more gas-poor and their gas fractions decrease as the host halo's mass increases.

\subsubsection{Gas content: stacked phase-space planes}
The examples above illustrate the diversity in dynamical states and surrounding environments for the clusters in our sample. In order to draw general conclusions about how halo gas richness varies with location in the phase-space plane, we next stack the data for all 324 clusters.



\Fig{fig:all_stack} shows the result of stacking the phase-space planes from all 324 clusters on a $50\times50$ grid. The two columns correspond to the stacked phase-space plane in two perspectives: the 6D (`true') is shown on the left-hand side and the LOS (`observed') projection is shown on the right-hand side. The details of how the phase-space planes were calculated and stacked in each projection are contained in \Sec{sec:phase-space}. 

The pink panels in the top row show the number of haloes and subhaloes in each bin. Here, the contours indicate lines of constant halo and subhalo numbers. The 6D perspective shows two distinct infalling and outgoing populations of objects. The population with ${\rm v}_{\rm 6D}/\sigma_{\rm 6D}<0$ are presumably made up of not only first-infalling objects, but also objects that have backsplashed and are making a subsequent infall. Even though there is some overlap between these two populations, the peaks are distinct enough that from this plot alone, one may satisfactorily disentangle the two populations. However, it is clear that the outgoing population with ${\rm v}_{\rm 6D}/\sigma_{\rm 6D}>0$ will be made up of mostly backsplash objects, hence why much of this population doesn't extend much further out than $\sim\,2\,{\rm R}_{\rm 200}$. These bimodal populations are totally washed out in the LOS projection, where the only information available is the projected distance from the cluster and the magnitude of the LOS velocity. The first-infalling, backsplash, and second infalling populations all mix and from the phase-space coordinates alone, it is impossible to disentangle these populations.

The middle panels show the median gas fraction of all haloes and subhaloes stacked in each bin, and the bottom panels show the standard deviation of the gas fraction distribution in each bin. We present the median gas fractions, as the gas fraction distribution is roughly reverse (skewed right) log-normal in each gas-rich pixel. The main result here is the cut-off in gas fraction at $\sim 1.5-2\ \text{R}_{\text{200}}$ in infalling objects. Similarly to \Fig{fig:sing_clus}, haloes and subhaloes seem to be losing nearly all of their gas at this radius on their first infall, but here we see the same effect is seen when averaged across several hundred clusters and over $>10^5$ haloes and subhaloes. The fall off in gas content at this radius suggests that objects have either been completely stripped of their gas or, alternatively, their gas has been transformed into stars and has not been replenished. However, if the objects have been stripped, it could suggest the presence of an accretion shock at $\sim 1.5-2\ \text{R}_{\text{200}}$, where objects start to experience hydrodynamical environmental effects before even reaching ${\rm R}_{\rm 200}$. This agrees with \citet{power2018} who modelled the spherical density, temperature and radial velocity profiles for the cluster gas surrounding a synthetic cluster, and from these found evidence to suggest the presence of what could be shocks due to infalling gas from the cosmic web colliding with a relatively hydrostatic halo of hot gas. 

In the LOS perspective the cut-off in object gas fraction occurs slightly further out, closer to $\sim\,2\,\text{R}_{\text{200}}$. Within this radius, haloes and subhaloes are drastically more gas-poor. We note that the shape of the gas-rich region is consistent with that found in fig. 3 of \citet{jaffe2015}. However, they find H\,{\sc i} detected galaxies as close to the cluster as  $\sim\,0.5\,\text{R}_{\text{200}}$. In our study we may also find such objects, but they may be averaged out by the large sample size in each bin on the phase-space plane. However, in contrast an alternate suggestion to this discrepancy is that our simulations have too poor resolution to resolve the cold gas discs of galaxies within the haloes and subhaloes, which likely results in numerically enhanced stripping in our models. Objects above $\sim\,1.25\,{\rm v}_{\rm LOS}/\sigma_{\rm LOS}$ are also frequently gas-poor well beyond $\text{R}_{\text{200}}$. These regions could be active sites on the phase-space plane where pre-processing is occurring, in the form of gas removal. These high speed regions in the outskirts in the LOS plane roughly correspond to the first infalling population in the 6D plane with more negative velocities than the main branch. As discussed previously with regards to the middle cluster in \Fig{fig:sing_clus}, the more negative velocities of these first-infallers is due to their host environment in the cluster outskirts. The haloes and subhaloes that reside at high speed regions in the outskirts in the LOS plane could potentially contain the red and dead galaxies that are unexplained by the model proposed in \citet{jaffe2015} and it is likely that they are being pre-processed in our simulations. 

From the bottom-left panel in \Fig{fig:all_stack}, it is clear that there is more gas fraction scatter from the infalling population of objects than the outgoing population. This shows that the infalling population are not losing their gas instantaneously at a certain radius, but gradually lose it as they infall. Individual objects may still be instantaneously stripped at different radii here, our results only show that the population as a whole is not being stripped instantaneously at the same radius. Despite this, the panel above still describes a characteristic radius ($\sim 1.5-2\ \text{R}_{\text{200}}$) where first infalling objects start to become heavily affected by their cluster environment. The dispersion reduces to zero at $\sim\,0.5\,\text{R}_{\text{200}}$ for the infalling population, and stays at this value for the outgoing population. In the LOS projection, the scatter is fairly constant across the plane, with the exception of objects at small cluster-centric radii and high relative velocities; here objects are more gas-poor. 

It has been suggested in \citet{jaffe2018} that the correlation between the gas content of cluster galaxies and their phase-space location has some dependence on host cluster mass. On the other hand though, \citet{rhee2017} found that the correlation with halo tidal mass loss and phase-space location is not linked to cluster mass, for clusters under $\lesssim10^{15}\ h^{-1}\text{M}_{\odot}$. We explore this further in \Fig{fig:clus_mass_bins}, by binning our stacked phase-space planes with respect to cluster halo mass, $\text{M}_{\text{200}}$. Here \threehun\ dataset allows us to expand this analysis well beyond $10^{15}\ h^{-1}\text{M}_{\odot}$ with a statistically significant cluster count in each mass bin. Each column represents one cluster mass bin, where the mass range is indicated above the top panel. The mass ranges of the bins have been chosen to contain 81 clusters each.

We see from the top panels in \Fig{fig:clus_mass_bins} that the number of haloes on the phase-space plane increases with respect to cluster mass. This is perhaps unsurprising, as one would expect a larger cluster to attract more material in the outskirts, as well as the centre. What is surprising is the lack of discrepancy between cluster mass bins in the middle-row panels. The phase-space planes showing the median gas fractions look extremely similar to one another. In fact, the difference between the average gas fractions in the lowest and highest cluster mass phase-space planes is only $\sim 0.015$. Again we see the sharp cut-off in gas-richness at $\sim 1.5\ \text{R}_{\text{200}}$ in every cluster mass bin. Though many of our clusters are massive, our cluster mass range still spans an order of magnitude. Therefore, one might expect to see some differences in the correlation between gas fraction and phase-space position between the bins. Our null result suggests that cluster mass has little bearing on when haloes and subhaloes lose their gas.

\subsubsection{Gas content: separating haloes and subhaloes}  
In addition to studying the phase-space plane dependency with cluster mass, we can use the same methods to study differences between haloes and subhaloes on these planes. \Fig{fig:sing_clus_gfrac_substat} showed the gas content differences between the halo and subhalo populations in the cluster 0005 example, but we have explored this further by stacking the phase-space planes for all 324 clusters, separating out haloes and subhaloes in \Fig{fig:all_stack_substat}.

From the top-left panel, we see that no haloes reside within ${\rm R}_{\rm 200}$, whereas in the top-right panel the sharp transition at this boundary delineates what \ahf\ defines as a cluster subhalo and a subhalo of an infalling object. Approximately $10$ percent of objects beyond ${\rm R}_{\rm 200}$ are subhaloes and some of these are outgoing objects, which may indicate that these subhaloes travelled through the cluster as part of another infalling backsplash object. 

The middle panels show that both haloes and subhaloes seem to lose their gas at $\sim 1.5\ \text{R}_{\text{200}}$. However, haloes are substantially more gas-rich than subhaloes beyond $\text{R}_{\text{200}}$. Of the haloes and subhaloes that still contain gas at $z=0$, subhaloes contain $\sim 80$ percent less gas than their halo counterparts, on average, in the outskirts. We note that there are less subhaloes in each bin, which makes these median gas fractions somewhat less reliable. However, we suggest that this is good evidence for pre-processing of subhaloes by infalling host environments, especially if one considers that infalling subhaloes with more negative velocities are definitively more gas-poor than the main branch of infalling subhaloes.

\subsection{Ram pressure}\label{sec:results_ramp}

\begin{figure}
    \centering
    \includegraphics[width=\columnwidth]{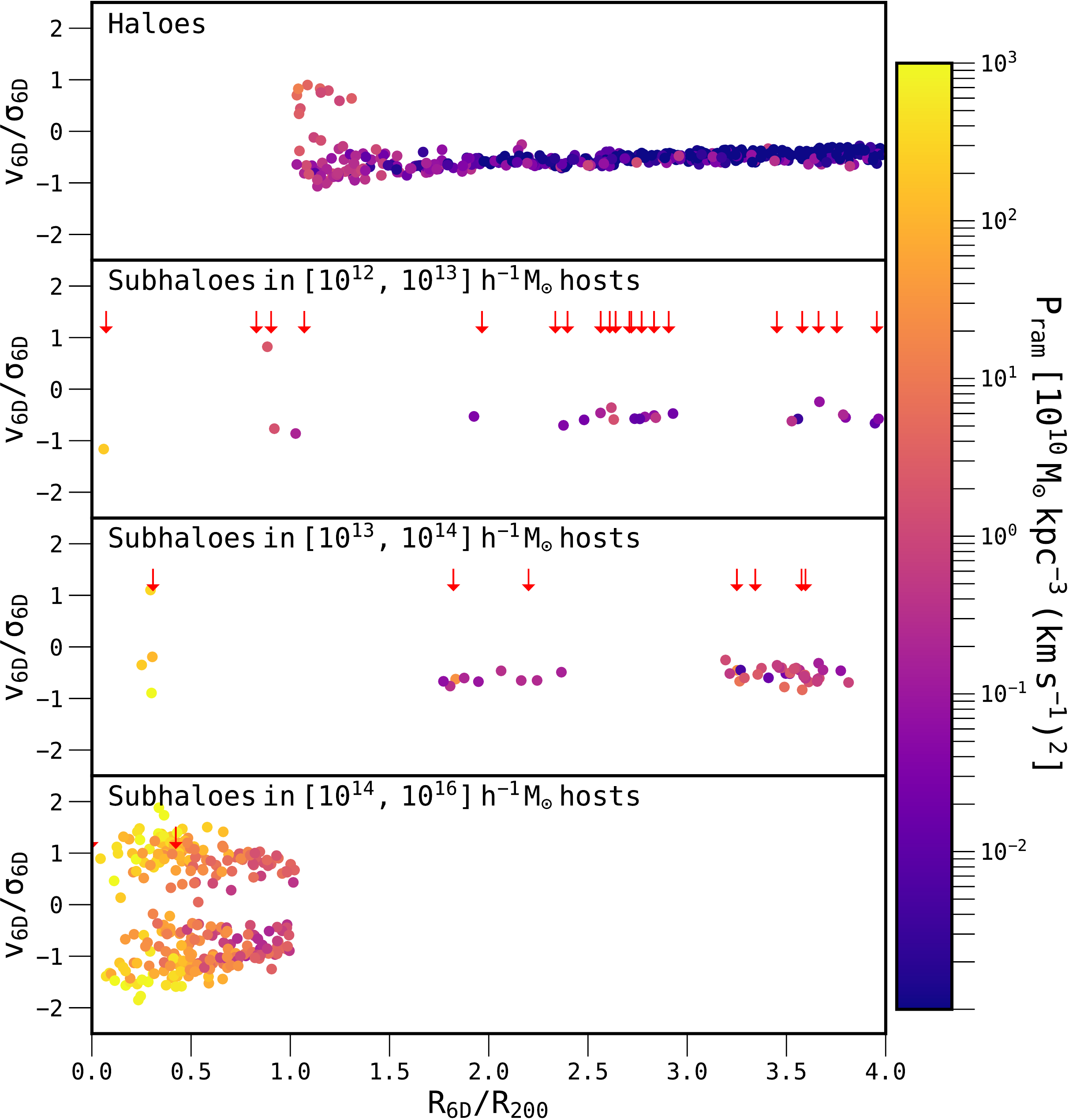}
    \caption{As in \Fig{fig:sing_clus_gfrac_substat}, but each halo and subhalo are now colour-coded by instantaneous ram pressure, calculated using the method described in \Sec{sec:ram pressure}. The red arrows in the bottom three  panels mark the positions of the host haloes that the subhaloes reside in. The figure demonstrates how we can use the wealth of information provided by \threehun\ to investigate how the instantaneous ram pressure correlates with phase-space position, cluster environment and subhalo status. In this example, subhaloes are experiencing more ram pressure compared to haloes at the same phase-space location and this seems to be proportional with their host mass.}
    \label{fig:sing_clus_ramp_substat}
\end{figure}

\begin{figure*}
    \centering
    \includegraphics[width=0.8\textwidth]{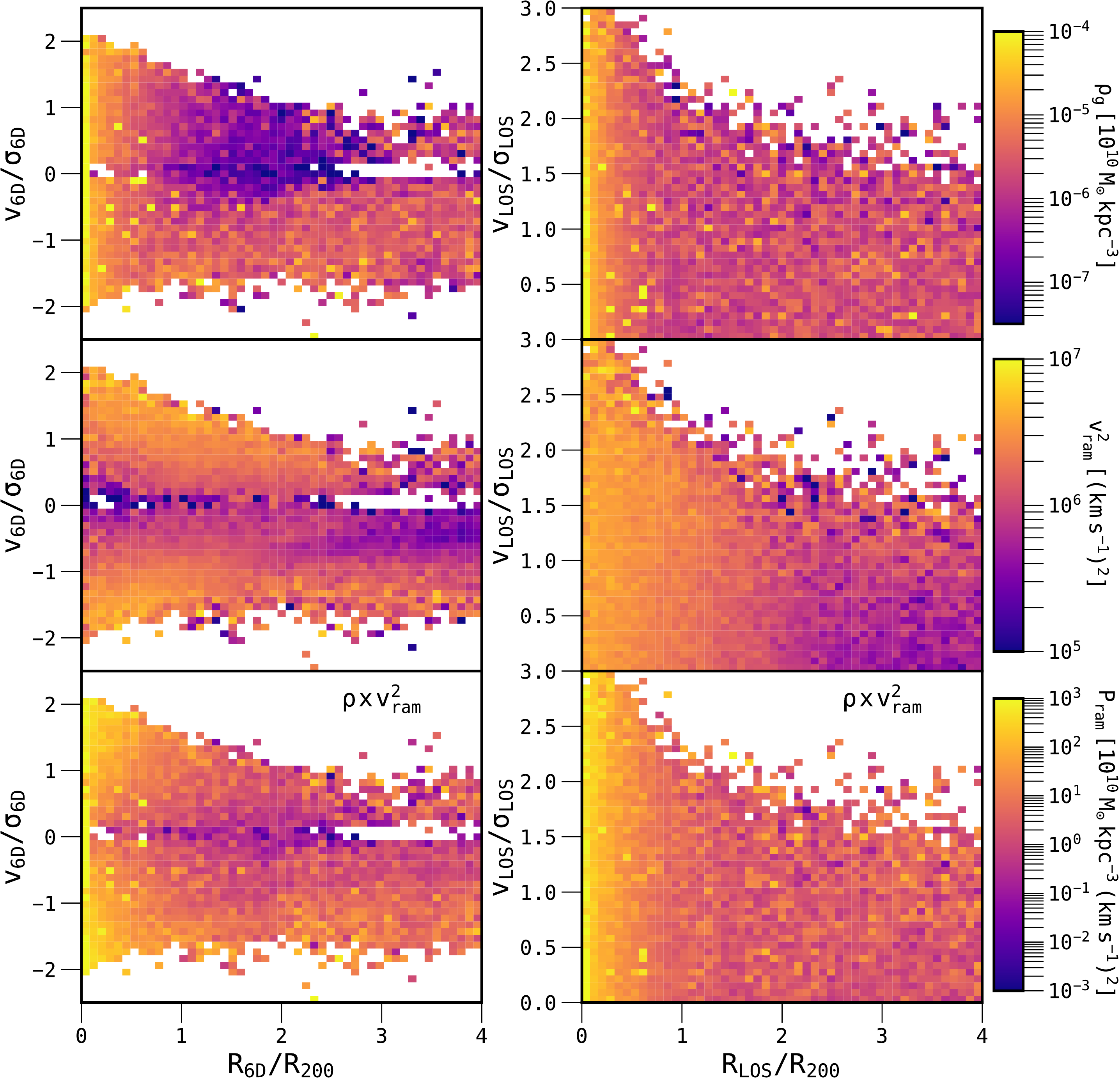}
    \caption{The stacked phase-space planes for all 324 clusters in the \threehun\ project, including both haloes and subhaloes. Here, the panels depict the effects of each term in \Eq{pram}, used for calculating the instantaneous ram pressure. The planes are stacked on a $50\times50$ grid where the top row is colour-coded by the median local gas density in each bin, the middle row is colour-coded by the median differential velocity between the objects in each bin and their local gas, and the bottom row is colour-coded by the product of the first two rows as calculated by using the method described in \Sec{sec:ram pressure}. The left and right columns show the 6D and LOS perspectives, respectively. The 6D perspective shows that infalling objects typically experience higher local gas density than the outgoing population, but this is washed out in the LOS perspective. There are distinct regions in the phase-space planes with low ${\rm v}_{\rm ram}^2$, and these line up well with the regions of high object gas fractions in \Fig{fig:all_stack}.}
    \label{fig:pps_all_stack_ramp}
\end{figure*}

\begin{figure*}
    \centering
    \includegraphics[width=0.7\textwidth]{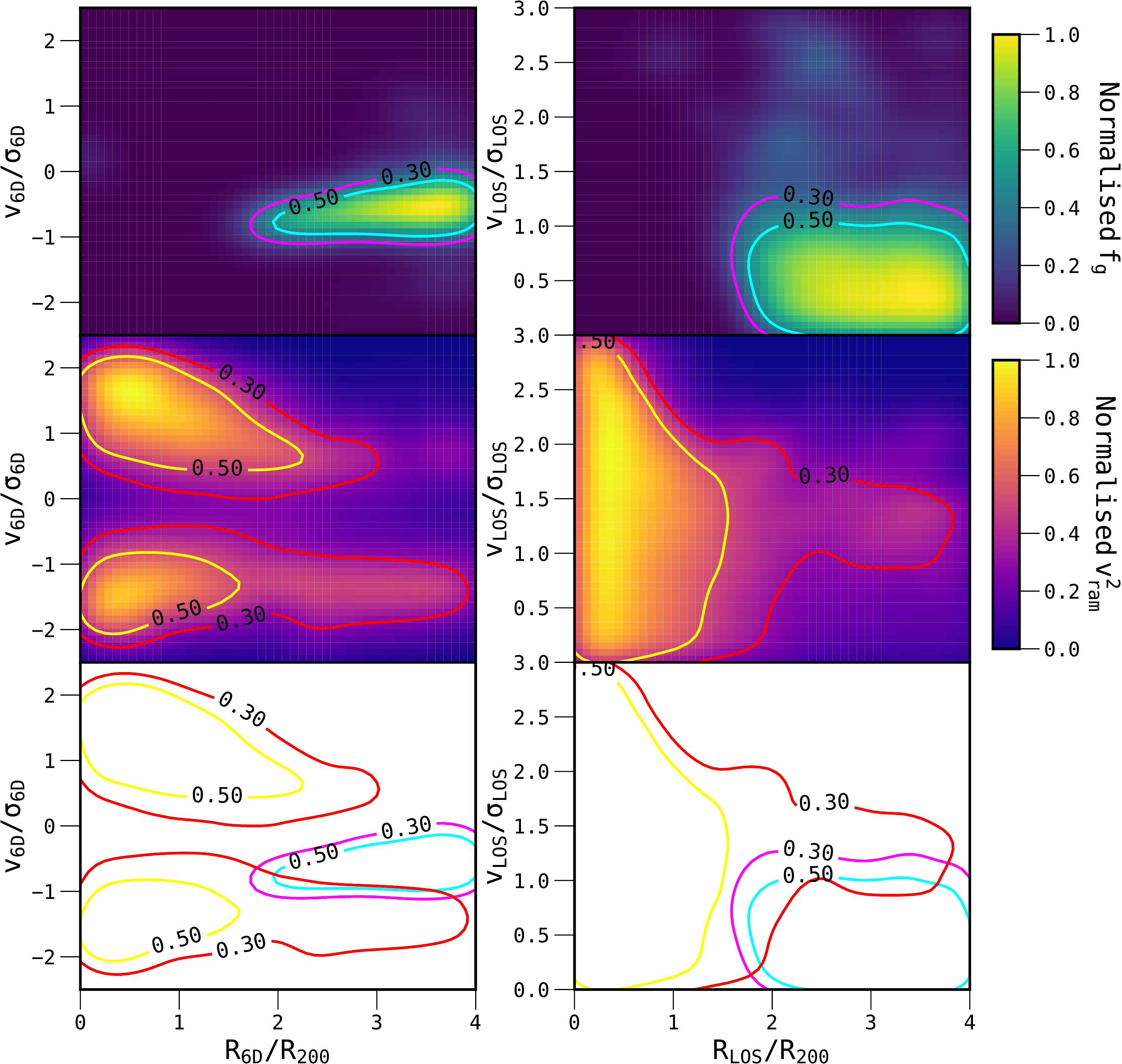}
    \caption{In both the 6D and LOS perspectives, the figure shows the middle-row panels from \Fig{fig:all_stack} (top row) and  \Fig{fig:pps_all_stack_ramp} (middle row). Here though, the planes have been convolved with a gaussian and then normalised. The bottom row illustrates how the countours from each plane align. Though there is some overlap, the contours from $\text{f}_{\text{g}}$ and ${\rm v}_{\rm ram}^2$ appear almost mutually exclusive with each other.}
    \label{fig:fg_ramp_contour}
\end{figure*}

\begin{figure}
    \centering
    \includegraphics[width=\columnwidth]{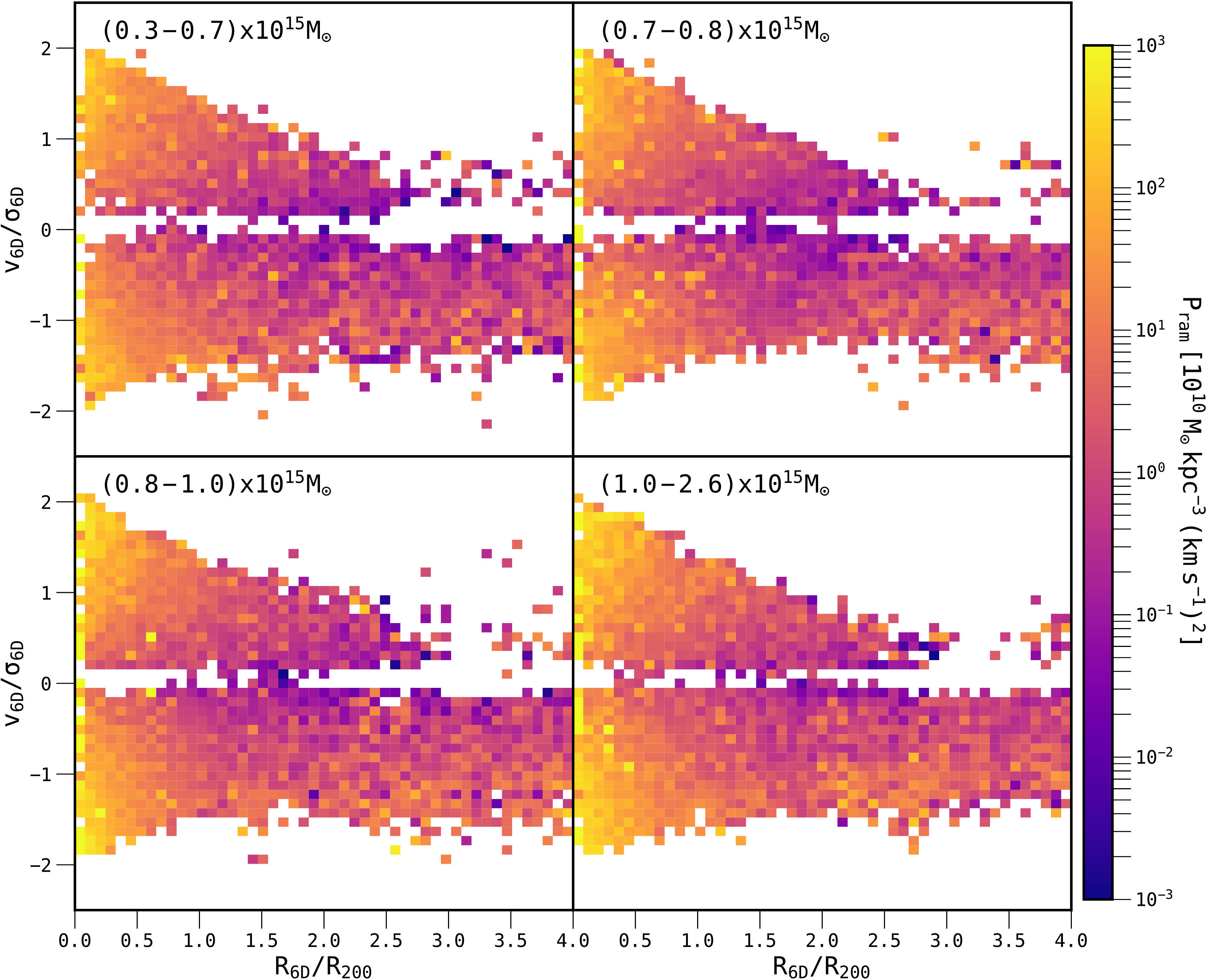}
    \caption{As in \Fig{fig:clus_mass_bins}, but coloured according to the median instantaneous ram pressure on haloes and subhaloes in each bin. Like in \Fig{fig:clus_mass_bins}, there appears to be no trend between cluster mass and instantaneous ram pressure in the phase-space plane in our cluster sample.}
    \label{fig:clus_mass_bins_ramp}
\end{figure}

\begin{figure}
    \centering
    \includegraphics[width=\columnwidth]{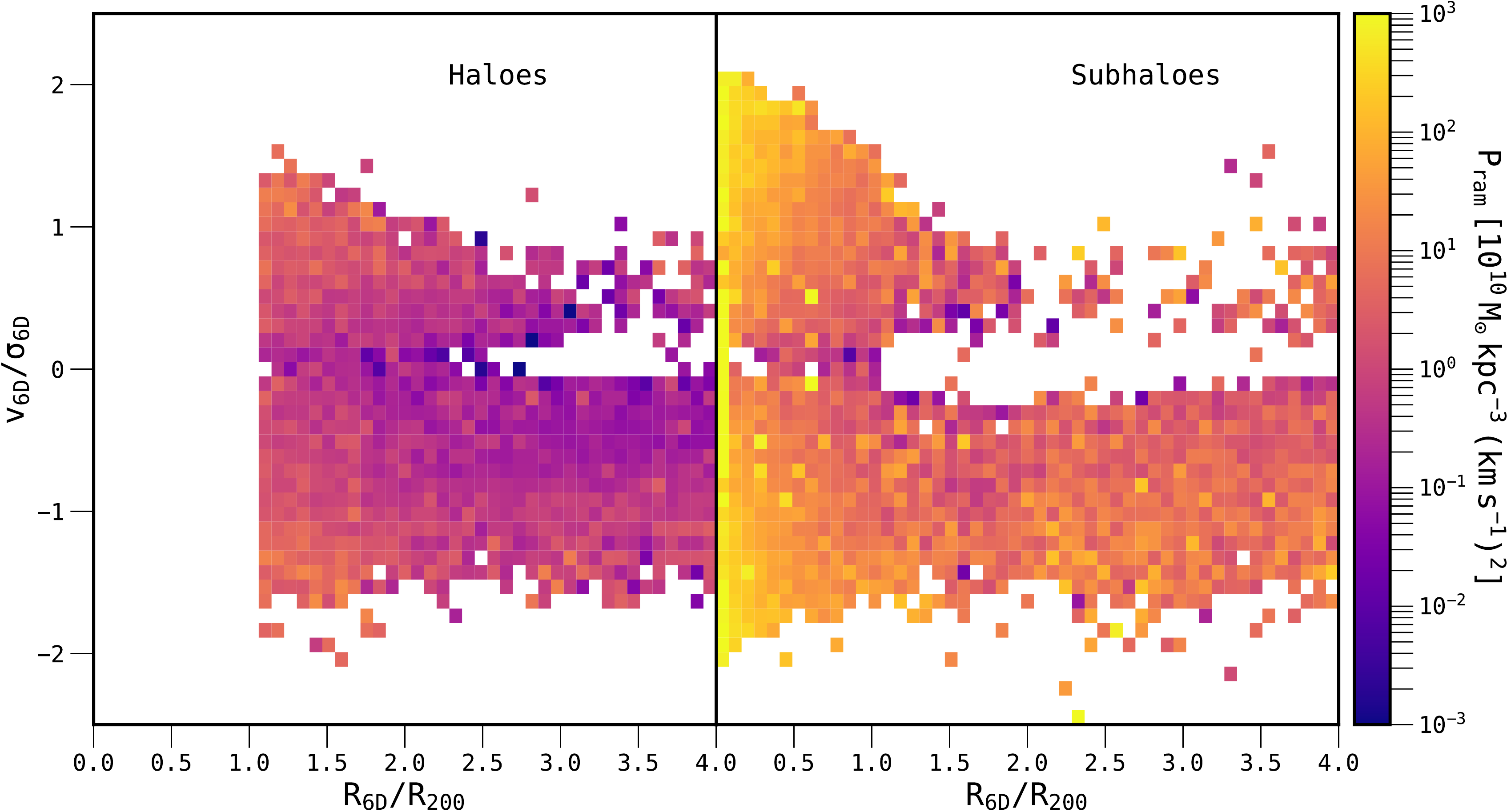}
    \caption{As in \Fig{fig:all_stack_substat}, but coloured according to the median instantaneous ram pressure on haloes and subhaloes in each bin. Subhaloes experience around two orders of magnitude more ram pressure than their halo counterparts in any region of the phase-space plane.}
    \label{fig:pps_all_stack_substat_ramp}
\end{figure}

In this section we investigate the link between the instantaneous ram pressure experienced by a halo or subhalo at $z=0$ and the gas content of these objects, using our phase-space planes. We build on the work of \citet{jaffe2015}, individually modelling the ram pressure incident on each halo using its local gas environment. Further details containing information about these calculations can be found in \Sec{sec:ram pressure}.

\subsubsection{Case study: a single massive cluster}
We start as we did in the previous section by first introducing the dataset via a single cluster. \Fig{fig:sing_clus_ramp_substat} shows the phase-space plane of the same cluster studied in \Fig{fig:sing_clus_gfrac_substat}, but here the figure is colour coded by the instantaneous ram pressure that each object experiences. The panels separate haloes and subhaloes in the same fashion as \Fig{fig:sing_clus_gfrac_substat}.

In this example, the fact that subhaloes are presently experiencing higher ram pressure than haloes is clear cut. The ram pressure they experience also seems to be correlated with the mass of their host. For example, subhaloes in hosts with mass $10^{13}\ h^{-1}\text{M}_{\odot}-10^{14}\ h^{-1}\text{M}_{\odot}$ on average experience a factor of $\sim 7.3$ more ram pressure than subhaloes in $10^{12}\ h^{-1}\text{M}_{\odot}-10^{13}\ h^{-1}\text{M}_{\odot}$ hosts. This effect is even more extreme in the bottom panel. As expected here for subhaloes in the cluster halo, there is a distinct trend with ram pressure and both distance from the centre of the cluster and velocity relative to the cluster.

\subsubsection{Ram pressure: stacked phase-space planes}
To explore these phenomena in more detail, we have stacked the phase-space planes of all 324 clusters in \Fig{fig:pps_all_stack_ramp}. Here we show the 6D and LOS perspective planes in the left- and right-hand columns, respectively. The bottom pair of panels show how the median instantaneous ram pressure intensity, calculated with \Eq{pram}, is correlated with phase-space position. It is clear that the trends we saw above for a single object are evident when we combine all our re-simulations together. In order to explore the influence of the density and velocity terms independently, we have also included the top and middle panels, which show how each variable in \Eq{pram} varies across the phase-space plane. The top panels show how the local gas density around all objects varies with phase-space position, whereas the middle panels indicate the same for the differential velocity between objects and their local gas.

Firstly, the top panels show that in the central region within $\sim 0.5\ \text{R}_{\text{200}}$, the local gas density around haloes and subhaloes is largely governed by the gas density profile of the cluster. Within this region, the local gas density has little relation with phase-space velocity, $\text{v}/\sigma$, and falls off with increasing cluster-centric distance. On average, it seems that the infalling population of haloes and subhaloes experience higher ram pressures than the outgoing population. This is an interesting result, as it suggests that infalling objects are being funnelled into the cluster as part of a dense environment, such as a group or filament, whilst the outgoing population passes through the cluster and is distributed more isotropically. This result isn't insignificant either, as between $\sim\,1-2.5\,{\rm R}_{\rm 200}$, on average the local gas density is $\sim\,2$ orders of magnitude higher in the infalling population. On the other hand, an alternate explanation may be that any loosely bound gas that has recently accreted onto an infalling object may be stripped by the time it has passed through the cluster. This could explain the discrepancy between the local gas density surrounding infalling and backsplash objects. A future study (Haggar et al. (in prep.)) will fully investigate the backsplash population around \threehun\ and could shed more light on which explanation is more plausible.

Similar results are found in the top-right panel in the LOS perspective within $\sim 0.5\ \text{R}_{\text{200}}$. Again, in this region the local gas density falls off with cluster-centric distance. Outside this region there appears to be no trend between phase-space position and local gas density. 

In the 6D phase-space planes, the differential velocity, ${\rm v}^2_{\rm ram}$, between an object's trajectory and their local gas medium is strongly correlated with phase-space coordinate. Firstly, and perhaps unsurprisingly, we find that the differential velocity is not as well correlated with cluster-centric distance, but rather with the relative phase-space velocity of an object. This is especially true within $\text{R}_{\text{200}}$, where the fastest moving objects have ${\rm v}^2_{\rm ram}$ over one order of magnitude more than the slower objects. This suggests that the gas in the cluster is somewhat stationary with respect to the cluster. 

There is a population of infalling objects at $>2\,\text{R}_{\text{200}}$, which have much lower differential velocities compared to other infalling objects, made up of mostly infalling haloes. We find a similar population of objects at high cluster-centric distance and low phase-space velocity that experience lower differential velocities in the LOS perspective. We find that the differential velocity planes closely mirror the gas fraction phase-space planes from the middle planes shown in \Fig{fig:all_stack}. This striking relationship holds in both perspectives and is explored further in  \Fig{fig:fg_ramp_contour}. 

The bottom panels of \Fig{fig:pps_all_stack_ramp} display the combination of the variables in \Eq{pram} from the above panels. The trends in the above panels are carried through here in some parts of the plane. Within $\sim\,0.5\,\text{R}_{\text{200}}$, the ram pressure is dominated by the local gas density, but a trend with relative phase-space velocity can also be seen due to the differential velocity term. Outside $\text{R}_{\text{200}}$, one can see how the lower density gas environments around the outgoing population contribute to lower ram pressure. However, in the cluster outskirts, the effects from the differential velocity are somewhat washed out by the local gas density. This is true in both perspectives. We note that within $\sim1\,\text{R}_{\text{200}}$, our LOS ram pressure planes are similar to those found in \citet{jaffe2015}. 

We have further investigated the relationship between an object's gas fraction and the differential velocity, ${\rm v}_{\rm ram}^2$, in \Fig{fig:fg_ramp_contour}. We show the middle panes from both \Fig{fig:all_stack} and \Fig{fig:pps_all_stack_ramp} in both perspectives. However, here we have convolved all planes with a gaussian and normalised each by their maximum value after smoothing. Isovalue contours have been added to each plot to show where $30$ percent and $50$ percent of each panel's maximum value lie. These contours are overlaid in the bottom panels. 

The contours illustrate how ${\rm f}_{\rm g}$ and ${\rm v}^2_{\rm ram}$ correlate with each other. What is noticeable at first glance is how mutually exclusive they appear. In regions of high gas fraction, one does not see high differential velocities, and vice versa. This is true in both the 6D and LOS perspectives. The mutual exclusivity of these two parameters suggests that they are causally linked, and that objects are primarily losing their gas via stripping, where the dominant factor is how fast an object is travelling with respect to its local gas environment, as expected from \Eq{pram}.  The contours and the gaussian kernel ($\sigma = 2$ pixels), used for the convolution, have been chosen arbitrarily in order to illustrate how ${\rm f}_{\rm g}$ and ${\rm v}^2_{\rm ram}$ align with each other. The encapsulated regions are somewhat sensitive to which contour values are chosen, but the two parameters are nearly always mutually exclusive for $\gtrsim\,30$ percent contours.

\subsubsection{Ram pressure: separating haloes and subhaloes}
We can subdivide the bottom two panels of \Fig{fig:pps_all_stack_ramp} with respect to cluster mass in a similar fashion to that used previously for \Fig{fig:clus_mass_bins}. The results are shown in \Fig{fig:clus_mass_bins_ramp}.

\Fig{fig:clus_mass_bins_ramp} demonstrates that the correlation between instantaneous ram pressure and phase-space location does not vary significantly with cluster mass. This corresponds to the relation between gas fraction and phase-space as a function of cluster mass shown in \Fig{fig:clus_mass_bins}, further suggesting that the gas content of (sub)haloes and ram pressure are causally linked. When averaged across the whole 6D plane, haloes in the most massive bin only experience a factor of $\sim 1.12$ more ram pressure than those in the lowest mass bin. As would be expected, any residual differences between these two planes appear to be more substantial within $\sim 1\ \text{R}_{\text{200}}$, rather than in the cluster outskirts.

Our results suggest that the correlation between phase-space coordinate and ram pressure has little dependence on cluster mass. However, as mentioned before, our sample does consist of very massive clusters. Our lowest mass cluster, with mass $\text{M}_{\text{200}}=\,6.08\ \text{x}\ 10^{14}\ h^{-1}\text{M}_{\odot}$, is still a large object. It would be interesting to see how our phase-space planes vary for not only less massive clusters, but also group environments ($\text{M}_{{\rm 200}} = 10^{13}-10^{14}\text{M}_{\odot}$) in the outskirts as well. A further study targeting these questions may enable us to infer how much of an influence group environments have on infalling galaxies, and could shed more light on pre-processing. Alongside this, by extending the mass range, such a study may help us infer how much of an effect the mass of a group/cluster has on infalling objects. This scaling with host mass is preliminarily seen in \Fig{fig:sing_clus_ramp_substat} for several host environments in the outskirts of our single cluster example.

We also investigate how the ram pressure intensity correlates with whether an object is a halo or a subhalo in \Fig{fig:pps_all_stack_substat_ramp}. It is immediately obvious that subhaloes experience higher ram pressures than haloes, and that this is true across the whole phase-space plane, even in the outskirts. Averaging (but not weighting by the number of objects) across the outskirts of the 6D plane, subhaloes experience around $\gtrsim 2$ orders of magnitude higher ram pressure. We even find subhaloes that experience ram pressures of order $10^{11} \! \lesssim \! {\rm P}_{\rm ram} / \left({\rm M}_{\odot}\,{\rm kpc}^{-3}\,{\rm km}^2\,{\rm s}^{-2}\right) \! \lesssim \! 10^{12}$, in the high-velocity infalling outskirts, which again indicates that these regions could be active sites for pre-processing by group environments. Subhaloes are presumably experiencing more ram pressure due to their halo environment, rather than any large scale structure feeding the clusters, such as filaments.

\section{Summary and Conclusions}\label{sec:conclusions}
 
We use \threehun\ project, a suite of 324 resimulated massive galaxy clusters embedded in a broad range of environments, to investigate how the gas content of surrounding haloes and subhaloes correlates with phase-space position at $z=0$. We have also studied how the instantaneous ram pressure experienced by these objects varies across the phase-space plane and what role it plays in the gas content of these objects at $z=0$. Our main conclusions may be summarised as follows:

\begin{itemize}
\item \threehun\ project is an ideal dataset to study environmental effects on the phase-space plane. The suite consists of 324 resimulated galaxy clusters, with masses between $\text{M}_{\text{200}}=6.08\ \text{x}\ 10^{14}\ h^{-1}\text{M}_{\odot}$ and $\text{M}_{\text{200}}=2.62\ \text{x}\ 10^{15}\ h^{-1}\text{M}_{\odot}$, at $z=0$. Each resimulation contains a full description of the dark matter, gas, and stars in a volume that extends well beyond $5\ \text{R}_{\text{200}}$ ($\text{R}_{\text{200}}$ is $\text{R}_{\text{200},\text{crit}}$ of the cluster halo in each resimulation), which also enables access to the detailed large-scale structure surrounding each cluster. 
\item By stacking all 324 cluster phase-space planes in both the 6D and Line-Of-Sight (LOS) perspectives, we have shown that the gas content of haloes and subhaloes is tightly correlated with phase-space position. There is a distinct cut-off in gas content of infalling objects at $\sim 1.5-2\ \text{R}_{\text{200}}$ in the 6D perspective, where objects are most likely to be on their first infall \citep{oman2013}. These single-snapshot results are in agreement with \citet{lotz2018}, who recently showed that objects are significantly altered on their first infall. We postulate that this sharp cut-off in gas content could indicate the presence of an accretion shock at $\sim 1.5-2\ \text{R}_{\text{200}}$, which is in agreement with what is found in \citet{power2018}. Projection effects blur this line somewhat, and make it appear as though objects lose their gas closer to $\sim 2\ \text{R}_{\text{200}}$. 
\item We investigate how the correlation between object gas content and phase-space position changes depending on cluster mass. This is an important question, as it sheds more light on what environmental effect the cluster halo has on infalling objects. Like in \citet{rhee2017}, who found that the correlation between halo tidal mass loss and phase-space position was not dependent on cluster mass, we found no dependence on cluster mass for halo gas content in our massive galaxy cluster sample. Irrespective of mass bin, the stacked phase-space planes show that infalling objects lose their gas at $\sim 1.5-2\ \text{R}_{\text{200}}$. A future study will investigate how the gas fraction of objects on the phase-space plane changes within groups.
\item Subhaloes are considerably more gas-poor than haloes. Although subhaloes are completely devoid of gas within ${\rm R}_{\rm 200}$ as expected, on average subhaloes in the outskirts contain $\sim 80$ percent less gas than haloes that are also in the outskirts. Therefore, this indicates that these subhaloes may be being pre-processed by their host environment before they even reach the cluster halo. 
\item From the stacked phase-space planes, we have shown that there are distinct regions on the phase-space plane where haloes and subhaloes experience greater instantaneous ram pressure as they fall in. By examining the local gas density around an object, we see that within ${\rm R}_{\rm 200}$, the local gas could be well modelled with a spherical gas density profile and is only really dependent on cluster-centric distance. However, outside of this region the local gas density seems to be more dependent on an object's phase-space velocity. Between $\sim 1-2.5\ \text{R}_{\text{200}}$, infalling objects are surrounded by $\sim\,2$ orders of magnitude more dense gas than outgoing objects. This is an interesting result, as it suggests that one of two processes could happen exclusively or inconjuction with each other. The first is that infalling objects are being funnelled into the cluster as part of a dense environment, such as a group or filament, whilst the outgoing population passes through the cluster and is distributed more isotropically. On the other hand, the other explanation may be that any loosely bound gas that has recently accreted onto an infalling object may be stripped by the time it has passed through the cluster. A future study (Haggar et al. (in prep.)) will further investigate the backsplash population and shed more light in this area.  
\item We have also shown how the differential velocity term between an object and its local gas varies across our stacked phase-space planes. There are distinct regions in the phase-space plane where objects experience high differential velocities, and these are almost mutually exclusive with areas of high gas fraction. This suggests that there is not only a causal link between an object's gas fraction and the instantaneous ram pressure it experiences, but that the differential velocity term is dominant in determining the gas content of an object. These findings are consistent with the literature \citep[e.g.][]{gunn1972}
\item Lastly, we find that the correlation between instantaneous ram pressure and phase-space coordinate has little dependence on cluster mass. Any differences between each mass bin are primarily located within $\sim\,1\,{\rm R}_{\rm 200}$.
\end{itemize}

We  have  utilised  the  power  of  a large suite of cluster re-simulations to examine the gas content of haloes and subhaloes as a function of position on the phase-space plane. Our findings indicate that ram pressure plays an important  role  in  determining  gas  content,  particularly close to the possible dynamic location of the cluster accretion shock that lies outside, but close to the cluster virial radius. The use of such a large suite of simulations has allowed us to examine whole-population effects and reach robust conclusions without the confounding factors of variation in substructure content and cluster dynamical state  that  can dominate for individual clusters  or small samples. A future study will be a temporal analogue to this paper. By using full halo orbital histories, we will track the whole halo and subhalo population to further investigate the interplay between gas content and ram pressure in the phase-space plane.

\section*{Acknowledgments}
We thank the anonymous referee for all their very useful comments and suggestions which has led to a greatly improved paper.

The work has received financial support from the European Union's Horizon 2020 Research and Innovation programme under the Marie Sklodowskaw-Curie grant agreement number 734374, i.e. the LACEGAL project\footnote{https://cordis.europa.eu/project/rcn/207630\_en.html}. The workshop where this work has been finished was sponsored in part by the Higgs Centre for Theoretical Physics at the University of Edinburgh.

The authors would like to thank The Red Espa\~nola de Supercomputaci\'on for granting us computing time at the MareNostrum Supercomputer of the BSC-CNS where most of the cluster simulations have been performed. The MDPL2 simulation has been performed at LRZ Munich within the project pr87yi. The CosmoSim database (\href{https://www.cosmosim.org}{https://www.cosmosim.org}) is a service by the Leibniz Institute for Astrophysics Potsdam (AIP). Part of the computations with \gx\ have also been performed at the `Leibniz-Rechenzentrum' with CPU time assigned to the Project `pr83li'.

JTA acknowledges support from a postgraduate award from STFC.
PJE is supported by the ARC CoE ASTRO 6D through project number CE170100013.
WC, AK and GY are supported by the {\it Ministerio de Econom\'ia y Competitividad} and the {\it Fondo Europeo de Desarrollo Regional} (MINECO/FEDER, UE) in Spain through grant AYA2015-63810-P. WC secondly acknowledges the supported by the European Research Council under grant number 670193. WC further thanks TaiLai Cui for all the joys. AK is also supported by the Spanish Red Consolider MultiDark FPA2017-90566-REDC and further thanks Ladytron for light and magic. CP acknowledges the Australia Research Council (ARC) Centre of Excellence (CoE) ASTRO 6D through project number CE170100013. The authors also gratefully acknowledge Giuseppe Murante.

\section*{Affiliations}
\noindent
{\it
$^{1}$School of Physics \& Astronomy, University of Nottingham, Nottingham NG7 2RD, UK\\
$^{2}$Departamento de F\'isica Te\'{o}rica, M\'{o}dulo 15 Universidad Aut\'{o}noma de Madrid, 28049 Madrid, Spain\\
$^{3}$Centro de Investigaci\'{o}n Avanzada en F\'{\i}sica Fundamental (CIAFF), Universidad Aut\'{o}noma de Madrid, 28049 Madrid, Spain \\
$^{4}$International Centre for Radio Astronomy Research, The University of Western Australia, 35 Stirling Highway, Crawley, Western Australia 6009, Australia\\
$^{5}$ Institute for Astronomy, University of Edinburgh, Royal Observatory, Edinburgh EH9 3HJ, United Kingdom\\
$^{6}$University Observatory Munich, Scheinerstra{\ss}e 1, 81679 Munich, Germany\\
$^{7}$Max-Planck-Institute for Extraterrestrial Physics, Giessenbachstrasse 1, 85748 Garching, Germany\\
$^{8}$Department of Physics, Sapienza Universita di Roma, p.le Aldo Moro 5, I-00185 Rome, Italy\\
$^{9}$INFN - Sezione di Roma, P.le A. Moro 2, I-00185 Roma, Italy\\
$^{10}$Max-Planck Institute for Astrophysics, Karl- Schwarzschild-Strabetae 1, D-85741 Garching, Germany\\
$^{11}$Department of Astronomy $\&$ Astrophysics, University of Toronto, Toronto, Canada\\
$^{12}$European Space Astronomy Centre (ESAC), Villanueva de la Canada, E-28692 Madrid, Spain\\
$^{13}$INAF - Osservatorio Astronomico di Trieste, via Tiepolo 11, I-34143 Trieste, Italy\\
}


\bibliographystyle{mnras}
\bibliography{phasespacez0} 

\bsp	
\label{lastpage}
\end{document}